\documentclass[sn-mathphys-num]{sn-jnl}

\usepackage{graphicx}%
\usepackage{multirow}%
\usepackage{amsmath,amssymb,amsfonts}%
\usepackage{amsthm}%
\usepackage{mathrsfs}%
\usepackage[title]{appendix}%
\usepackage{xcolor}%
\usepackage{textcomp}%
\usepackage{manyfoot}%
\usepackage{booktabs}%
\usepackage{algorithm}%
\usepackage{algorithmicx}%
\usepackage{algpseudocode}%
\usepackage{listings}%

\theoremstyle{thmstyleone}%
\newtheorem{theorem}{Theorem}%

\theoremstyle{thmstyletwo}%

\theoremstyle{thmstylethree}%
\newtheorem{definition}{Definition}%

\begin{document}

\title[Article Title]{Non-Hermitian tearing by dissipation}

\author[1]{\fnm{Qian} \sur{Du}}

\author[2]{\fnm{Xin-Ran} \sur{Ma}}

\author*[2]{\fnm{Su-Peng} \sur{Kou}}\email{spkou@bnu.edu.cn}

\affil[1]{\orgdiv{School of Physics and Electronic Engineering}, \orgname{Linyi University}, \orgaddress{\city{Linyi}, \postcode{ 276000}, \country{China}}}

\affil[2]{\orgdiv{Center for Advanced Quantum Studies, Department of Physics}, \orgname{Beijing Normal
University}, \orgaddress{\city{Beijing}, \postcode{100875},  \country{China}}}

\abstract{In the paper, we study the non-Hermitian system under dissipation and give the
effective $2\times2$ Hamiltonian in the $k$-space by reducing the $N\times N$
Hamiltonian in the real space for them. It is discovered that the energy band
shows an imaginary line gap. To describe these phenomena, we propose the
theory of \textquotedblleft non-Hermitian tearing\textquotedblright%
,\thinspace \ in which the tearability we define reveals a continuous phase
transition at the exceptional point. The non-Hermitian tearing manifests in
two forms --- separation of bulk state and decoupling of boundary state. In
addition, we also explore the one-dimensional Su-Schrieffer-Heeger model and
the Qi-Wu-Zhang model under dissipation using the theory of non-Hermitian tearing. Our results provide a theoretical
approach for exploring the controlling of non-Hermitian physics on
topological quantum states.}


\maketitle

\section{Introduction}\label{sec1}

Non-Hermitian systems have been a hot topic owing to their unique properties
and potential applications in various fields, such as optics
\cite{C2010,A2009,Y2011,R2018,L2013,H2014,L2014,W2019}, condensed matter
physics \cite{V2017,Z2018,H2018,F2019,M2020,K2019,K2021,Y2020}, and quantum
mechanics \cite{Bender02,I2009,F2012,M2002}. In the fundamental principles of
quantum mechanics, the physical quantity describing the state of a microscopic
system is a Hermitian operator in Hilbert space, whose expected value is a
real number. However, in practice, we find that the probability of a system
does not always conserve, and the eigenvalues of energy can also be complex
\cite{N2011,U2020,Bender07}. Thus, non-Hermitian operators become crucial. The
origins of the non-Hermitian Hamiltonian can be traced back to the lifetime of
a quasiparticle \cite{G1928,P1982,R1971}, columnar defects in the
superconductor \cite{H1996}, and so on. Gradually, people's understanding of
quantum mechanics extended from Hermitian systems to non-Hermitian systems.

In recent years, non-Hermitian physics has witnessed remarkable advancement.
Bender pointed out in 1998 that the energy spectrum of a Hamiltonian
satisfying parity-time ($\mathcal{PT}$) symmetry can be classified into three
cases: all real numbers, complex conjugate pairs, and a situation with
spectral degeneracy and eigenstates merging, which is known as $\mathcal{PT}%
$-symmetry spontaneously broken \cite{Bender98}. This discovery has inspired
an array of theoretical and experimental breakthroughs in non-Hermitian
physics, including the non-Hermitian skin effect
\cite{V2018,Y2018,L2020,K2020,D2020,N2020} and the breakdown of bulk-boundary
correspondence \cite{D2020,X2018,T2016,F2018,S2018,H2021,X2020}. There has
been a great deal of research focused on non-Hermitian systems with global
non-Hermitian terms, such as gain and loss \cite{H2019,Y2022,J2022,D2023} and
nonreciprocal hopping \cite{D2023,S2023,T2019,J2023}, or the local non-Hermitian
terms \cite{C2023,B2022}. In particular, Ref. \cite{H2019} demonstrated the
arbitrary, robust light steering in reconfigurable non-Hermitian gain--loss
junctions by projecting the designed spatial pumping patterns onto the
photonic topological lattice. Ref. \cite{Y2022} studied photonic topological
insulators with different types of gain-loss domain walls and proposed an
effective Hamiltonian describing localized states and the corresponding
energies occurring at the domain walls. However, there are still many problems
worth studying.
For example, how exactly do the bulk states change with non-Hermitian terms? How do the boundary states appear and change?
Previous studies have mainly focused on the occurrence of boundary states with gain and loss, while the physical properties of other states were not considered.
In this paper, we not only study the changes of boundary states, but also the properties of bulk states.
We systematically investigate the properties of bulk states and boundary states under dissipation, especially
the characteristics of boundary states after bulk states are separated by gain and loss.

In the paper, we study an arbitrary one-dimensional tight binding model under
dissipation where the left and right sites are subject to different imaginary
potentials and give a series of effective $2\times2$ Hamiltonians
$h_{\mathrm{eff}}\left(  k\right)  $ in the $k$-space by reducing the $N\times
N$ Hamiltonian in the real space to understand it. It is shown that the
original single band is divided into two energy bands and then appears an
imaginary line gap. To describe these novel phenomena, we propose the concept
of \textquotedblleft non-Hermitian tearing\textquotedblright, in which the
system is either in the partial tearing or in the complete tearing. During these
processes, the energy eigenvalues display a $\mathcal{PT}$ transition.
Furthermore, we define the tearability to characterize the effect of different
imaginary potentials on an eigenstate. According to the relationship between the
direction of wave vectors and the interface's direction, we introduce two
types of non-Hermitian tearing --- separation and decoupling. Using the theory
of non-Hermitian tearing, we explore the physical properties of a simple
one-dimensional tight binding model, the one-dimensional Su-Schrieffer-Heeger
(SSH) model and the Qi-Wu-Zhang (QWZ) model. The results indicated that the
tearability exhibits a continuous phase transition at the exceptional point.
Bulk states show separation and boundary states show decoupling. We give
the effective Hamiltonian of bulk states and boundary states for them, which
fits well with numerical solutions. 
Our study contributes a theoretical approach to studying physical properties in more complex non-Hermitian systems.

The outline of this paper is as follows. In Sec. \ref{sec2}, based on an arbitrary
one-dimensional tight binding model with the imaginary potential, we give a
series of effective $2\times2$ Hamiltonians in the $k$-space by reducing the $N\times N$ Hamiltonian in the real space
to understand it. From this, we propose the theory of non-Hermitian tearing.
In Sec. \ref{sec3}, we take a simple one-dimensional tight binding model as an
example to explore the physical properties of non-topological systems with the
imaginary potential by the theory of non-Hermitian tearing. In Sec. \ref{sec4}, we
study the non-Hermitian tearing in the one-dimensional SSH model and discover
that two pairs of boundary states appear after bulk states are separated.
Moreover, these boundary states also show non-Hermitian tearing with a
$\mathcal{PT}$ transition. In Sec. \ref{sec5}, we discuss the same issue in the QWZ
model. In Sec. \ref{sec6}, we draw the conclusion.

\section{Non-Hermitian tearing}\label{sec2}

We consider an arbitrary one-dimensional tight binding model $H_{0}$ with a
negative imaginary potential $-iv$ on the left $N/2$ sites and a positive
imaginary potential $iv$ on the right $N/2$ sites under the periodic boundary
condition. The adjacent part between different imaginary potentials is named
the interface. To better understand the physical phenomena of the total
system, we give the effective Hamiltonian $H_{\mathrm{eff}}$ in the $k$-space by reducing the $N\times N$ Hamiltonian
in the real space, which is composed of a series of effective $2\times2$ Hamiltonians $h_{\mathrm{eff}}\left(  k\right)$.

\begin{theorem}[The effective Hamiltonian]\label{thm1}
The effective $2\times2$ Hamiltonian
$h_{\mathrm{eff}}\left(  k\right)  $ can be written as
\begin{equation}\label{eq1}
h_{\mathrm{eff}}\left(  k\right)  =\left(
\begin{array}
[c]{cc}%
h_{0}\left(  k\right)  -iv & \frac{\alpha \left(  k\right)  }{\sqrt{N_{1}}}\\
\frac{\alpha \left(  k\right)  }{\sqrt{N_{1}}} & h_{0}\left(  k\right)  +iv
\end{array}
\right)
\end{equation}
for $k=2\pi j/N_{1}$, $j=1,2,\cdots,N_{1}$
and $N_{1}=N/2$, where $h_{0}\left(  k\right)  $ is the
Hamiltonian of $H_{0}$ in the $k$-space and $v$ is
the imaginary potential strength. $\alpha \left(  k\right)  $
represents the coupling term at the interface, which is a $k$-dependent real number.
\end{theorem}

The eigenvalue of the effective Hamiltonian is
\begin{equation}
E_{\mathrm{eff}}\left(  k\right)  =E_{0}\left(  k\right)  \pm \sqrt
{\frac{\alpha^{2}\left(  k\right)  }{N_{1}}-v^{2}},
\end{equation}
where $E_{0}\left(  k\right)  $ is the eigenvalue of $h_{0}\left(  k\right)
$. Obviously, there is the $\mathcal{PT}$ transition. For a given $k$, as
$\alpha^{2}\left(  k\right)  >N_{1}v^{2}$, $E_{\mathrm{eff}}\left(  k\right)
$ is all real and the eigenvalues are in the phase with $\mathcal{PT}%
$-symmetry; as $\alpha^{2}\left(  k\right)  =N_{1}v^{2}$, $E_{\mathrm{eff}%
}\left(  k\right)  =E_{0}\left(  k\right)  $, the eigenvalues occurs the
$\mathcal{PT}$-symmetric spontaneous breaking; as $\alpha^{2}\left(  k\right)
<N_{1}v^{2}$, $E_{\mathrm{eff}}\left(  k\right)  $ is all complex and the
eigenvalues in the phase with $\mathcal{PT}$-symmetry breaking. With
increasing $v$, the eigenvalues transition from all real to all complex. In
the case of $\alpha^{2}\left(  k\right)  \ll N_{1}v^{2}$ for all $k$, all
eigenvalues are complex and the original single band $E_{0}\left(  k\right)  $
is divided into one energy band with a negative imaginary part $-i\sqrt
{v^{2}-\frac{\alpha^{2}\left(  k\right)  }{N_{1}}}$ and one energy band with a
positive imaginary part $i\sqrt{v^{2}-\frac{\alpha^{2}\left(  k\right)
}{N_{1}}}$. An imaginary line gap $\Delta=2i\sqrt{v^{2}-\frac{\alpha
^{2}\left(  k\right)  }{N_{1}}}$ appears between these two energy bands. In
order to describe the phenomenon, we give the following definitions:

\begin{definition}[Non-Hermitian tearing]
If the original single band of the system is
gradually divided into two energy bands due to the imaginary
potential, a down energy band $E_{\mathrm{down}}$ and an up energy
band $E_{\mathrm{up}}$, then an imaginary line energy gap emerges
between these two energy bands, %
\begin{equation}
\Delta=\min \left(  \operatorname{Im}E_{\mathrm{up}}\right)  -\max \left(
\operatorname{Im}E_{\mathrm{down}}\right)  .
\end{equation}
The phenomenon is called \textquotedblleft non-Hermitian
tearing\textquotedblright.
\end{definition}

In terms of the presence or absence of the imaginary line energy gap, we give \emph{partial tearing }and\emph{ complete tearing}.

\begin{definition}[Partial tearing and complete tearing]
If all energy eigenvalues are complex and the
imaginary line energy gap is present, $\Delta>0$, then we think that
the system is in the \textquotedblleft complete tearing\textquotedblright. If
some energy eigenvalues are still real and the imaginary line energy gap is
absent, $\Delta=0$, then we think that the system is in the
\textquotedblleft partial tearing\textquotedblright.
\end{definition}

In particular, when the system crosses over the critical point between the
partial tearing and complete tearing, the energy spectrum shows a
$\mathcal{PT}$ transition.

For one of these eigenvalues, to characterize the effect of different
imaginary potentials on its corresponding eigenstate, we introduce the
\emph{tearability}:

\begin{definition}[Tearability]
The ratio of the probability of an eigenstate %
$\Psi_{j}=\left(  \psi_{1},\cdots,\psi_{n}\cdots,\psi_{N}\right)  ^{\dag}%
$ in the right $N/2$ sites and left $N/2$ sites is
defined as the tearability of this eigenstate%
\begin{equation}
t_{j}=\frac{\rho_{j,\mathrm{R}}}{\rho_{j,\mathrm{L}}},
\end{equation}
where
\begin{equation}
\rho_{j,\mathrm{L}}=\sum_{n=1}^{N/2}|\psi_{n}|^{2},\text{ \  \ }\rho
_{j,\mathrm{R}}=\sum_{n=N/2+1}^{N}|\psi_{n}|^{2}.
\end{equation}
$\Psi_{j}$ is the right basis of Hamiltonian of the total system.
Thus, $\rho_{j,\mathrm{L}}+\rho_{j,\mathrm{R}}=1$.
\end{definition}
$t_{j}=1$ means that the probability in the right $N/2$ sites is the same as the probability in the left $N/2$ sites and the eigenstate is not torn. On the contrary, $t_{j}\neq1$ means that the probability in the right $N/2$ sites are different from the probability in the left $N/2$ sites and the eigenstate is torn. $t_{j}\rightarrow0$ or $t_{j}\rightarrow \infty$ means that
the corresponding wave function is bound in the left $N/2$ sites or
right $N/2$\ sites and the eigenstate is strongly torn.

In the non-Hermitian tearing, we think that the interface is directional. Its
direction in the one-dimensional tight binding model is identified as the
$y$-direction. We assume that the wave function corresponding to the
eigenstate is a plane wave and $k$ is the wave vector. According to the
relationship between the interface's direction and the direction of $k$, we
define $\emph{separation}$ and $\mathit{\emph{decoupling}}$:

\begin{definition}[Separation and decoupling]
If the direction of the wave vector of the eigenstate is
perpendicular to the interface's direction, then we call such a non-Hermitian
tearing \textquotedblleft separation\textquotedblright \ and the wave vector is
denoted by $k_{\perp}$. Conversely, if the direction of the wave vector of
the eigenstate is parallel to the interface's direction, then we call such a
non-Hermitian tearing \textquotedblleft decoupling\textquotedblright \ and the
wave vector is denoted by $k_{//}$.
\end{definition}

Given a more general case, a complex potential is applied to the
two-dimensional non-topological model
\begin{equation}
V    =v_{1}e^{i\phi_{1}}\sum_{n=1}^{N_{1}}c_{n}^{\dag}c_{n}+v_{2}e^{i\phi
_{2}}\sum_{n=N_{1}+1}^{N_{1}+N_{2}}c_{n}^{\dag}c_{n}
  +\cdots+v_{m}e^{i\phi_{m}}\sum_{n=N_{1}+\cdots+N_{m-1}+1}^{N_{1}+\cdots+N_{m-1}+N_{m}}%
c_{n}^{\dag}c_{n},
\end{equation}
where the real number $v_{1,2,\cdots,m}$ represents the magnitudes of the $m$
different complex potentials and $\phi_{1,2,\cdots,m}$ is the corresponding phase angle
with $\phi_{1,2,\cdots,m}\in \left[  0,2\pi \right]  $. Here, $N=N_{1}%
+N_{2}+\cdots+N_{m}$. In the system, the complex potential $v_{1}e^{i\phi_{1}%
}$ is applied to the $N_{1}$ sites, the complex potential $v_{2}e^{i\phi_{2}}$
is applied to the $N_{2}$ sites, and so on. Using a two-dimensional square
lattice model as an example, we consider the complex potential as shown in
Fig. 1(a), and show the complex energy spectrum in Fig. 1(b). The energy
spectrum of the system is torn to different positions along different
directions, depending on the form of the complex potential. Explicitly, the
direction is the phase angle of the complex potential, and the position is
relevant to the amplitude of the complex potential.

\begin{figure}[ptb]
\centering\includegraphics[clip,width=0.8\textwidth]{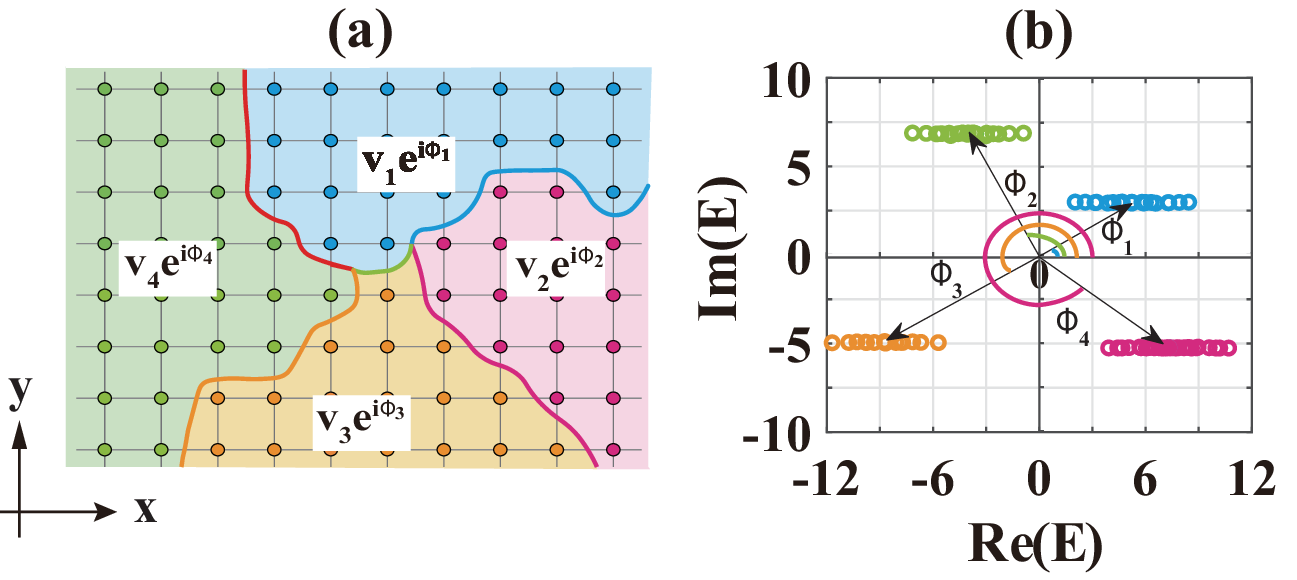}\caption{The two-dimensional
square lattice model with uniform hopping under the periodic boundary
conditions along the x-direction and y-direction. The four different complex
potentials are $v_{1}e^{i\phi_{1}}$, $v_{2}e^{i\phi_{2}}$, $v_{3}e^{i\phi_{3}%
}$, and $v_{4}e^{i\phi_{4}}$, respectively. (a) The model diagram. (b) The
energy spectrum. Here, $v_{1}=6$, $v_{2}=8$, $v_{3}=10$, $v_{4}=9$, and
$\phi_{1}=\frac{\pi}{6}$, $\phi_{2}=\frac{2\pi}{3}$, $\phi_{3}=\frac{7\pi}{6}%
$, $\phi_{4}=\frac{9\pi}{5}$.}%
\end{figure}

In the following sections, we will explore the physical properties of the
simple one-dimensional tight binding model, the one-dimensional SSH model, and
the QWZ model with the imaginary potential based on the theory of
non-Hermitian tearing.

\section{The simple one-dimensional tight binding model with the imaginary
potential}\label{sec3}
In this part, we mainly discuss the simple one-dimensional tight binding model
with the imaginary potential by the theory of non-Hermitian tearing to
investigate the effect of the imaginary potential on non-topological systems.

The Hamiltonian of the simple one-dimensional tight binding model in real
space is
\begin{equation}
H_{0}=t_{1}\sum_{n=1}^{N}c_{n}^{\dag}c_{n+1}+h.c.,
\end{equation}
where $t_{1}$ represents the hopping amplitude of an electron jumping from
site $n$ to site $n+1$. $c_{n}^{\dag}$ and $c_{n}$ are the creation and
annihilation operators of electron at site $n$, respectively. The Hamiltonian
in the $k$-space is
\begin{equation}
h_{0}\left(  k\right)  =2t_{1}\cos k,
\end{equation}
and its eigenvalue is%
\begin{equation}
E_{0}=2t_{1}\cos k.
\end{equation}
We consider the imaginary potential
\begin{equation}
V=-iv\sum_{n=1}^{N/2}c_{n}^{\dag}c_{n}+iv\sum_{n=N/2+1}^{N}c_{n}^{\dag}c_{n}%
\end{equation}
for the model, as shown in Fig. 2. In the paper, we use $t_{1}$ as the unit.
\begin{figure}[ptb]
\centering\includegraphics[clip,width=0.75\textwidth]{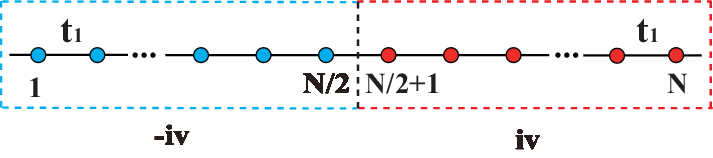}\caption{Illustration of the
simple one-dimensional tight binding model with the negative potential $-iv$
on the left $N_{1}=N/2$ sites and the positive potential $iv$ on the remaining
sites. $t_{1}$ represents hopping amplitude.}%
\end{figure}

Figure 3 plots the system's complex energy spectra and eigenstates for
different imaginary potential strength $v$ under the periodic boundary
condition. One can see that as $v$ goes up, the eigenvalues gradually move
along the positive or negative direction of the imaginary axis. Upon reaching
a particular imaginary potential, one energy band of the system is divided
into two energy bands: an up band $E_{\mathrm{up}}$ with a positive imaginary
part and a down band $E_{\mathrm{down}}$ with a negative imaginary part, as
shown in Fig. 3(c). As a consequence, an imaginary line gap appears between the
two energy bands. This indicates that the simple one-dimensional tight binding model with the
imaginary potential shows non-Hermitian tearing. In particular, there is a
$\mathcal{PT}$ transition in which the system transitions from partial tearing
to complete tearing in Fig. 3(b). Blue circles (red circles) represent the
energy eigenvalues moving along the negative (positive) direction of the
imaginary axis, and black circles represent unmoved energy eigenvalues. The
wave functions of moved energy eigenvalues (depicted by the blue curve or the
red curve) are bound to the left $N/2$ sites or the right $N/2$ sites, while
the wave functions of unmoved energy eigenvalues (depicted by the black curve)
still spread across all sites. In Fig. 3(a), some energy
eigenvalues are still real and the imaginary line energy gap is absent, so the
system is in the partial tearing. In Fig. 3(c), all energy
eigenvalues are complex and the imaginary line energy gap is present, so the
system is in the complete tearing. Note that these energy eigenstates are bulk
states and the directions of their wave vectors are perpendicular to the
interface's direction, so bulk states show separation.
\begin{figure}[ptb]
\centering\includegraphics[clip,width=0.8\textwidth]{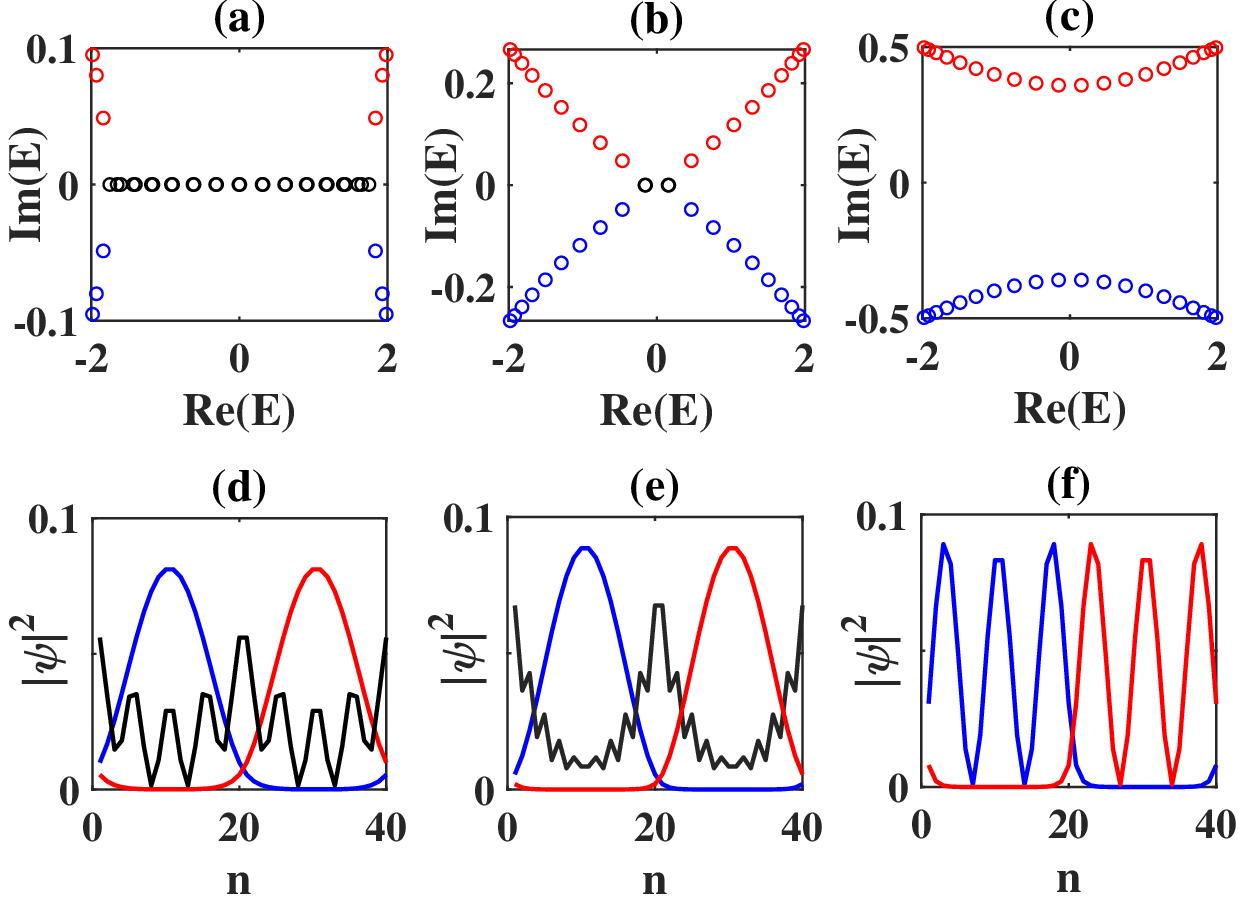}\caption{(a), (b) and (c)
are complex energy spectra for different imaginary potential strengths $v=0.1$,
$0.2695$ and $0.5$, respectively. (d), (e) and (f) are the corresponding bulk
states versus site $n$, respectively. The red, black or blue curve is a
representative wave function of energy eigenvalues represented by red, black
or blue circles. Here, $t_{1}=1$ and $N=40$. }%
\end{figure}

Considering that the energy eigenvalues move along the imaginary axis, we
arrange it in ascending order by imaginary part, $E_{1},\cdots,E_{j}%
,\cdots,E_{N}$, and show the corresponding probability in Fig. 4. The
probabilities of energy eigenvalues moving along the negative (positive)
direction of the imaginary axis are $\rho_{\mathrm{L}}>\rho_{\mathrm{R}%
}\left(  \rho_{\mathrm{L}}<\rho_{\mathrm{R}}\right)  $, whereas the
probabilities of unmoved energy eigenvalues are $\rho_{\mathrm{L}}%
=\rho_{\mathrm{R}}=0.5$. Together with Figs. 3(b) and 3(c), we can find that
in the partial tearing, there are some eigenstates with $t=1$ that correspond to
unmoved energy eigenvalues and are not torn. In the complete tearing, all
eigenstates have $t\neq1$ and are torn at the left $N/2$ sites or the right $N/2$
sites.
\begin{figure}[ptb]
\centering\includegraphics[clip,width=0.8\textwidth]{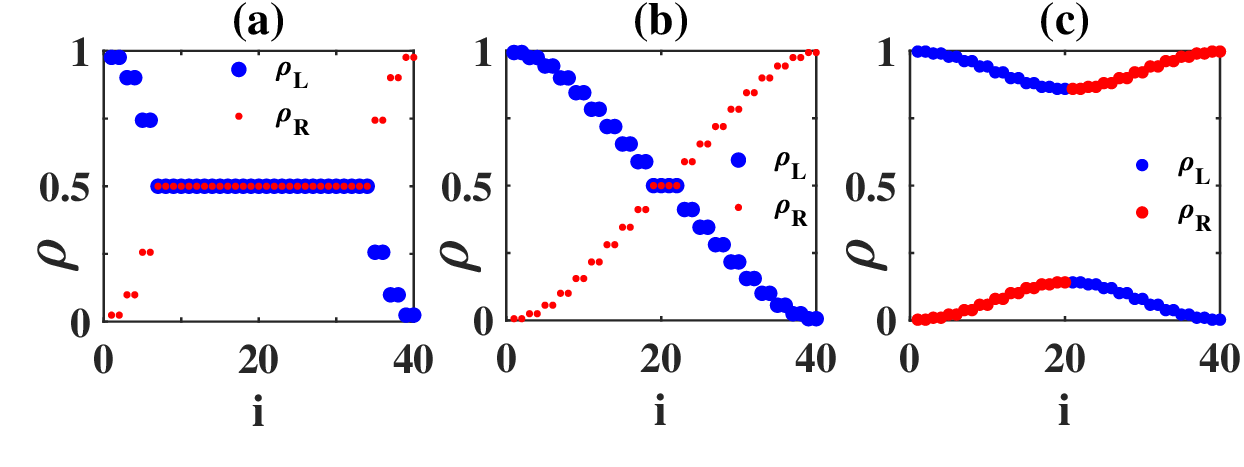}\caption{The probability of
the simple one-dimensional tight binding model with the imaginary potential, where
$i$ is the sort number. (a), (b) and (c) are the probabilities for different
imaginary potential strengths $v=0.1$, $0.2695$ and $0.5$, respectively. Here,
$t_{1}=1$ and $N=40$. }%
\end{figure}

Further, we calculate the tearability $t$ of the $j=11$ and the $j=31$ bulk
states in Figs. 5(a) and (c), respectively. The tearability $t=1$ indicates
that the bulk state is in the phase with $\mathcal{PT}$ symmetry and the
tearability $t\rightarrow0$ or $t\rightarrow \infty$ indicates that the bulk
state is in the phase with $\mathcal{PT}$-symmetry breaking. They are
continuous at the exceptional point where their energy eigenvalues transition
from real numbers to complex numbers. Later on, we calculate their derivatives
$\frac{\partial t}{\partial v}$ in Figs. 5(b) and (d). $\frac{\partial
t}{\partial v}$ is discontinued at the exceptional point, which means a
second-order phase transition at the exceptional point.
\begin{figure}[ptb]
\centering\includegraphics[clip,width=0.7\textwidth]{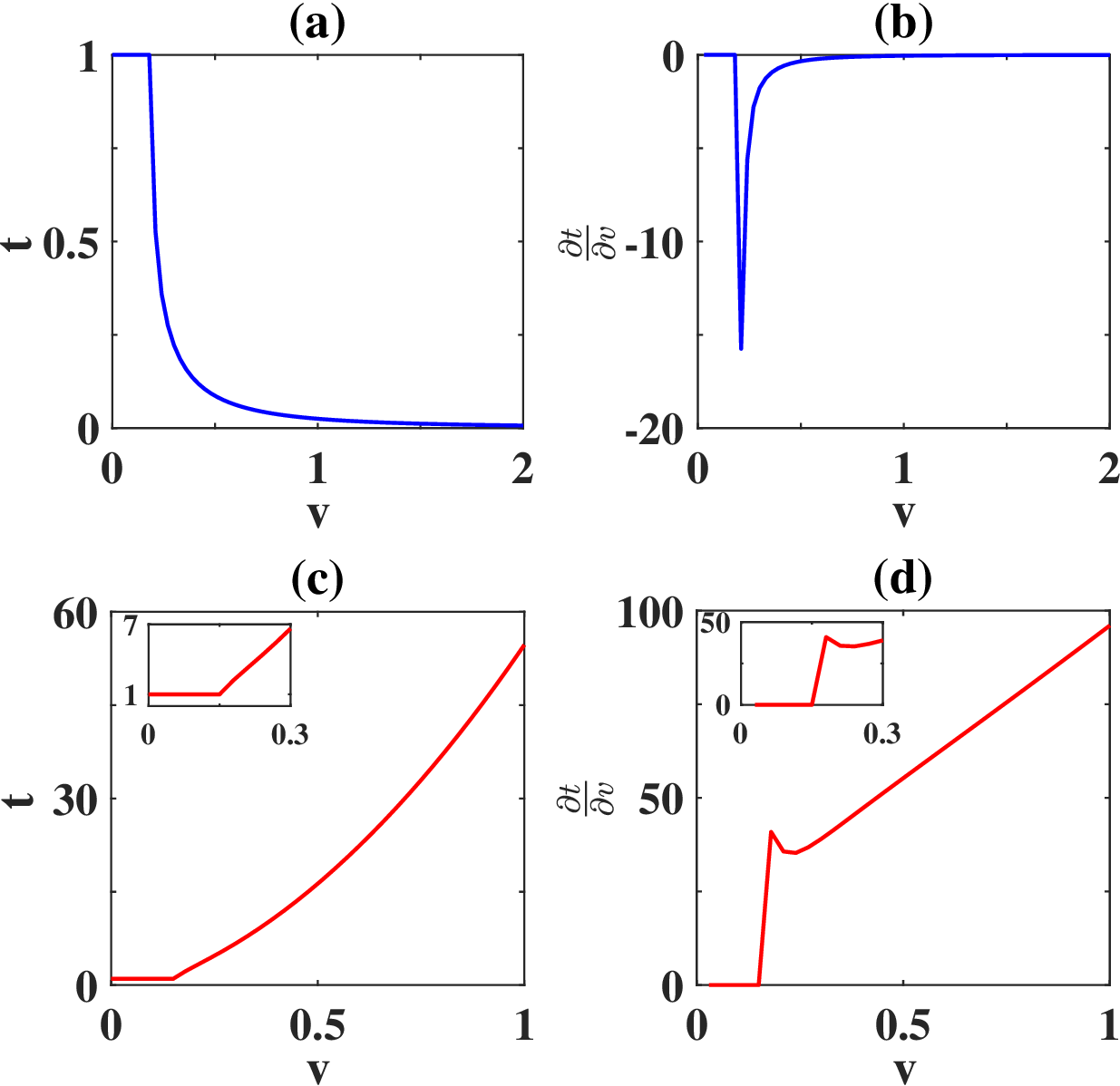}\caption{(a) and (b) are the
tearability $t$ and the derivative $\frac{\partial t}{\partial v}$ of the
$j=11$ bulk state, respectively. (c) and (d) are the tearability $t$ and
derivative $\frac{\partial t}{\partial v}$ of the $j=31$ bulk state,
respectively. Here, $t_{1}=1$ and $N=40$.}%
\end{figure}

According to Theorem \ref{thm1} in Eq. (\ref{eq1}), we give the effective Hamiltonian
$h_{\mathrm{eff}}\left(  k\right)  $ of bulk states
\begin{equation}
\label{eq11}
h_{\mathrm{eff}}\left(  k\right)  =\left(
\begin{array}
[c]{cc}%
h_{0}\left(  k\right)  -iv & \frac{\alpha \left(  k\right)  }{\sqrt{N_{1}}}\\
\frac{\alpha \left(  k\right)  }{\sqrt{N_{1}}} & h_{0}\left(  k\right)  +iv
\end{array}
\right)  ,
\end{equation}
where $\alpha \left(  k\right)  =h_{0}\left(  k+\frac{3\pi}{2}\right)
\cdot \lambda=2t_1\sin k/\lambda$ and $\lambda$ is a fitting parameter related to $v$. Here, $k=2\pi j/N_{1}$, $j=1,2,\cdots,N_{1}$ and $N_1=N/2$.
We plot the complex energy spectrum from analytical solutions of the effective Hamiltonian
$H_{\mathrm{eff}}$ in the $k$-space and numerical solutions
of the Hamiltonian $H=H_{0}+V$ in the real space in Fig. 6(a). It can be seen
that the effective Hamiltonian fits well with numerical solutions,
demonstrating that the Theorem \ref{thm1} we give can well describe the properties of
one-dimensional tight binding models with the imaginary potential. See the
detailed calculations regarding the probability and tearability of eigenstates
of $h_{\mathrm{eff}}\left(  k\right)  $ in Appendix A. The imaginary line gap
is
\begin{equation}
\Delta=2\sqrt{\frac{4t_{1}^{2}\lambda^{2}\sin^{2}k}{N_{1}}-v^{2}}.
\end{equation}
As $\frac{4t_{1}^{2}\lambda^{2}\sin^{2}k}{N_{1}}-v^{2}=0$, namely,
\begin{equation}
k=k_{0}=\pm \arcsin \frac{v\sqrt{N_{1}}}{2t_{1}\lambda},
\end{equation}
we have $E_{\mathrm{eff}}\left(  k_{0}\right)  =E_{0}\left(  k_{0}\right)  $
and the bulk state is at the $\mathcal{PT}$ transition. Besides, we show the
fitting parameter $\lambda$ as a function of $v$ in Fig. 6(b). It can be seen that when $v$ is
increased to a large value, $\lambda$ reaches a saturation value $1$.
\begin{figure}[ptb]
\centering\includegraphics[clip,width=0.7\textwidth]{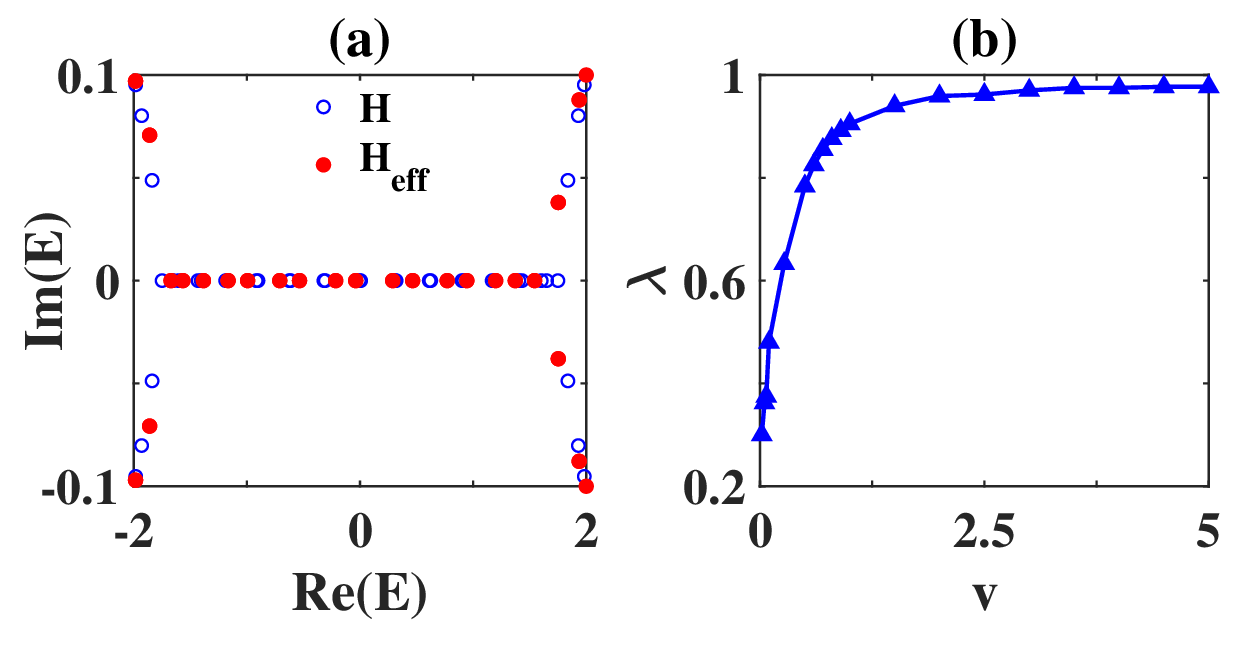}\caption{(a) The complex
energy spectrum from numerical solutions of the Hamiltonian $H$ in the real
space and analytical solutions of the effective Hamiltonian $H_{\mathrm{eff}}$
in the $k$-space, where $v=0.1$ and $\lambda=0.493$. (b) The fitting parameter
$\lambda$ versus the imaginary potential strength $v$. Here, $t_{1}=1$ and
$N=40$.}%
\end{figure}
In particular, if this model with the real potential, then the eigenvalues are all real and it does not show the $\mathcal{PT}$ transition from Eq. (\ref{eq11}).
When the system is torn, the tearability does not have a second-order continuous phase transition and is only at a crossover.

\section{The one-dimensional Su-Schrieffer-Heeger model with the imaginary
potential}\label{sec4}

In this section, we take the one-dimensional SSH model as an example and
explore the physical properties of one-dimensional topological systems with
imaginary potential.

The Hamiltonian of the one-dimensional SSH model in the real space is
\begin{equation}
H_{\mathrm{SSH}}=t_{1}\sum_{n=1}^{N}|n,B\rangle \langle n,A|+t_{2}\sum
_{n=1}^{N-1}|n+1,A\rangle \langle n,B|+h.c.,
\end{equation}
where $A$ and $B$ denote the two sublattices of each pair of lattice sites.
$t_{1}$ and $t_{2}$ describe the intra-cell and inter-cell hopping strengths,
respectively. Here, we use $t_{1}$ as the unit and set $t_{2}=2t_{1}=2$. The
Hamiltonian in $k$-space is
\begin{equation}
h_{\mathrm{SSH}}\left(  k\right)  =\left(  t_{1}+t_{2}\cos k\right)
\sigma_{x}+\left(  t_{2}\sin k\right)  \sigma_{y},
\end{equation}
where $\sigma_{i}$'s are the Pauli matrices acting on the sublattice subspace.
Its eigenvalue in the $k$-space is
\begin{equation}
E_{\pm}\left(  k\right)  =\pm \sqrt{\left(  t_{1}+t_{2}\cos k\right)
^{2}+\left(  t_{2}\sin k\right)  ^{2}}.
\end{equation}
Here, we consider the imaginary potential
\begin{equation}
V    =-iv\sum_{n=1}^{N/2}\left(  |n,A\rangle \langle n,A|+|n,B\rangle \langle
n,B|\right)
  +iv\sum_{n=N/2+1}^{N}\left(  |n,A\rangle \langle n,A|+|n,B\rangle \langle
n,B|\right)
\end{equation}
for the model as shown in Fig. 7.
\begin{figure}[ptb]
\centering\includegraphics[clip,width=0.8\textwidth]{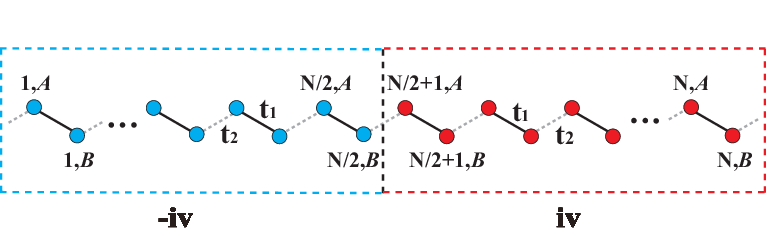}\caption{The one-dimensional
Su-Schrieffer-Heeger model with the negative potential $-iv$ on the left $N/2$
pairs of lattice sites and the positive potential $iv$ on the right $N/2$
pairs of lattice sites. $t_{1}$ and $t_{2}$ describe the intra-cell and
inter-cell hopping strengths, respectively.}%
\end{figure}

\begin{figure}[ptb]
\centering\includegraphics[clip,width=0.8\textwidth]{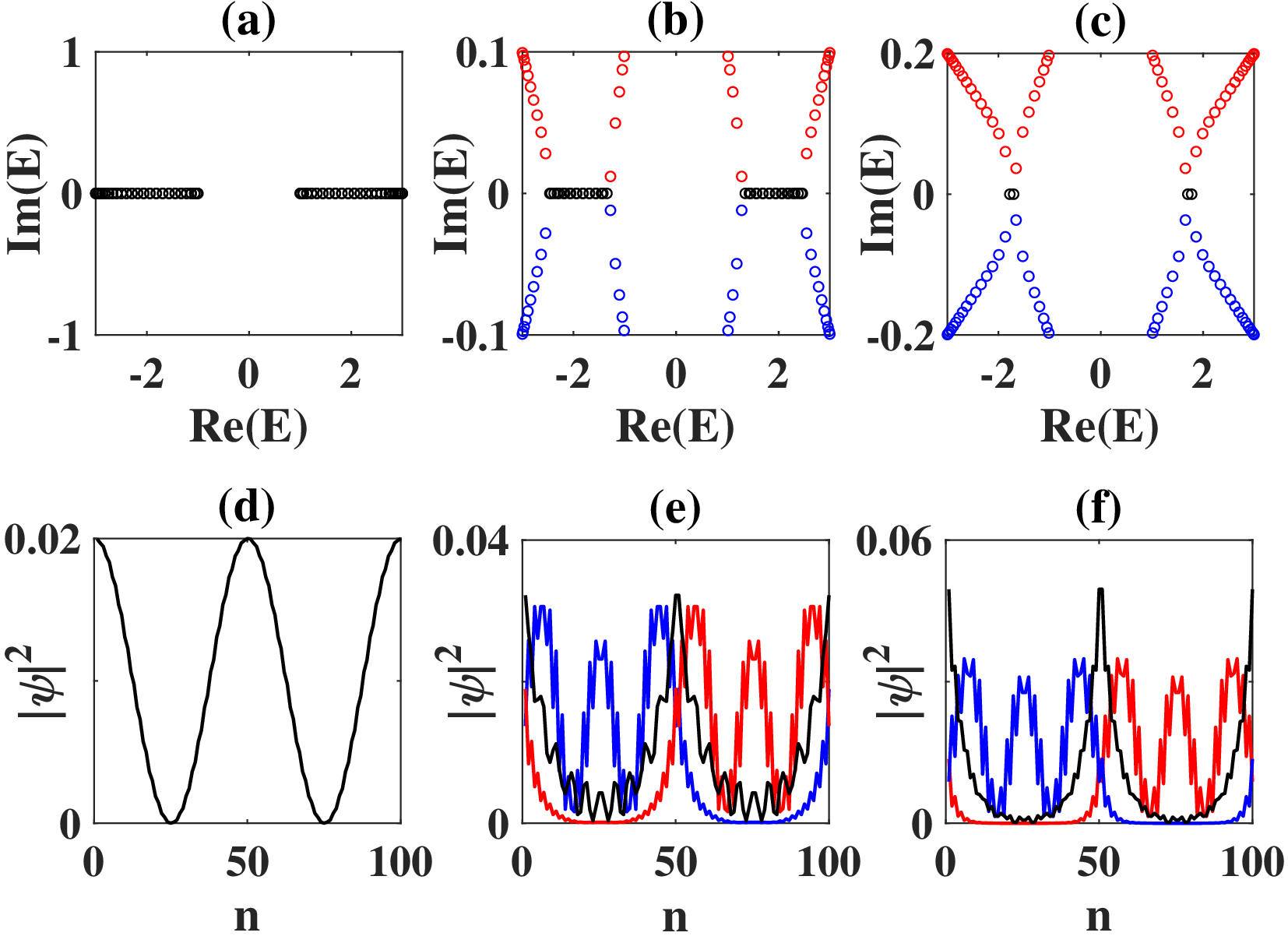}\caption{(a), (b) and (c)
are the complex energy spectra for different imaginary potential strengths $v=0,0.1$
and $0.2$, respectively. (d) to (f) are the corresponding bulk states versus
site $n$. The red, black or blue curve is a representative wave function of
energy eigenvalues represented by red, black or blue circles. Here,
$t_{1}=1,t_{2}=2,$ and $N=50$.}%
\end{figure}
Figures 8 and 9 plot the complex energy spectra and
eigenstates for different imaginary potential strengths $v$ under the periodic
boundary condition. With increasing $v$, each energy band of the system is
gradually divided into two energy bands: an up band $E_{\mathrm{up}}$ with a
positive imaginary part and a down band $E_{\mathrm{down}}$ with a negative
imaginary part, as shown in Fig. 9. An imaginary line gap emerges between the
two energy bands, meaning that the one-dimensional SSH model with the
imaginary potential shows non-Hermitian tearing accompanied by the
$\mathcal{PT}$ transition. In addition, bulk states show separation.
\begin{figure}[ptb]
\centering\includegraphics[clip,width=0.8\textwidth]{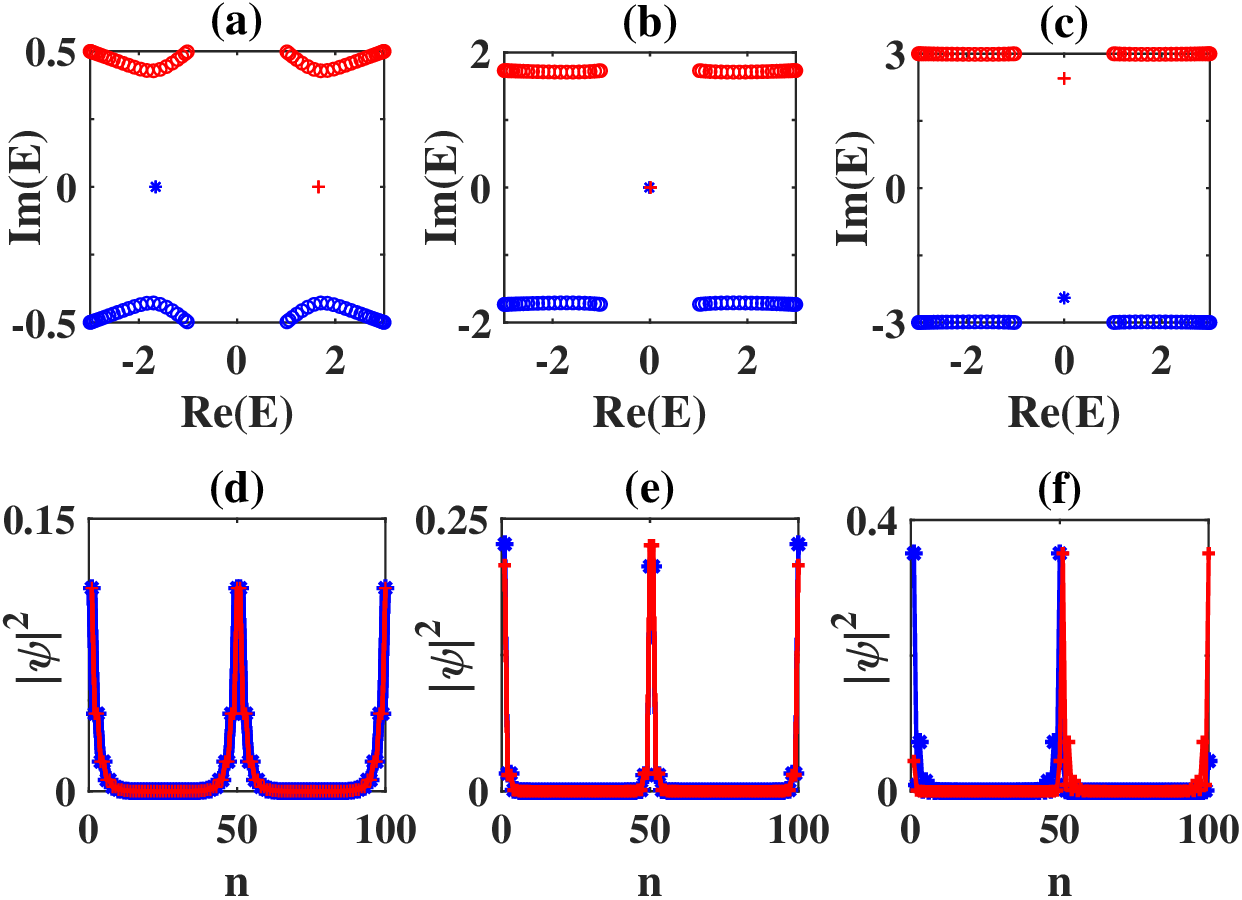}\caption{(a), (b) and (c)
are the complex energy spectra for different imaginary potential strengths
$v=0.5,1.732$ and $3$, respectively. (d), (e) and (f) are one of the left
(depicted by a blue curve with asterisks) and right (depicted by a red curve with
plus signs) boundary states, respectively. Here, $t_{1}=1,t_{2}=2,$ and
$N=50$.}%
\end{figure}

Surprisingly, after the bulk states are separated, the system appears two
pairs of boundary states as depicted in Fig. 9. As the imaginary potential
strength $v$ grows further, these two pairs of boundary states also show
non-Hermitian tearing along with a $\mathcal{PT}$ transition. Eventually, the
wave function of the boundary state corresponding to the down energy band
(represented by blue asterisks), which is depicted by the blue curve with
asterisks, is localized on the two boundaries of the left $N/2$ pairs of
lattice sites in Fig. 9(f). The wave function of the boundary state
corresponding to the up energy band (represented by red plus signs), which is
depicted by the red curve with red plus signs, is localized on the two
boundaries of the right $N/2$ pairs of lattice sites. The directions of wave
vectors of these boundary states are parallel to the interface's direction, so
these boundary states show decoupling.

To further investigate the non-Hermitian tearing of boundary states, we select
one of these two pairs of boundary states and calculate the tearability $t$
as a function of the imaginary potential strength $v$ in Figs. 10(a) and (c),
respectively. As $t=1$, the boundary state is not torn and its energy
eigenvalue is in the phase with $\mathcal{PT}$-symmetry, whereas as
$t\rightarrow0$ or $t\rightarrow \infty$, the boundary state is torn and its
energy eigenvalue in the phase with $\mathcal{PT}$-symmetry breaking. The
tearability is continuous at the exceptional point where their energy
eigenvalues transition from real numbers to imaginary numbers. Later on, we display
their derivatives $\frac{\partial t}{\partial v}$ versus $v$ in Figs. 10(b) and (d). It is evident that $\frac{\partial t}{\partial v}$ is discontinued
at the exceptional point, which implies that tearability has a second-order
phase transition at the exceptional point.
\begin{figure}[ptb]
\centering\includegraphics[clip,width=0.7\textwidth]{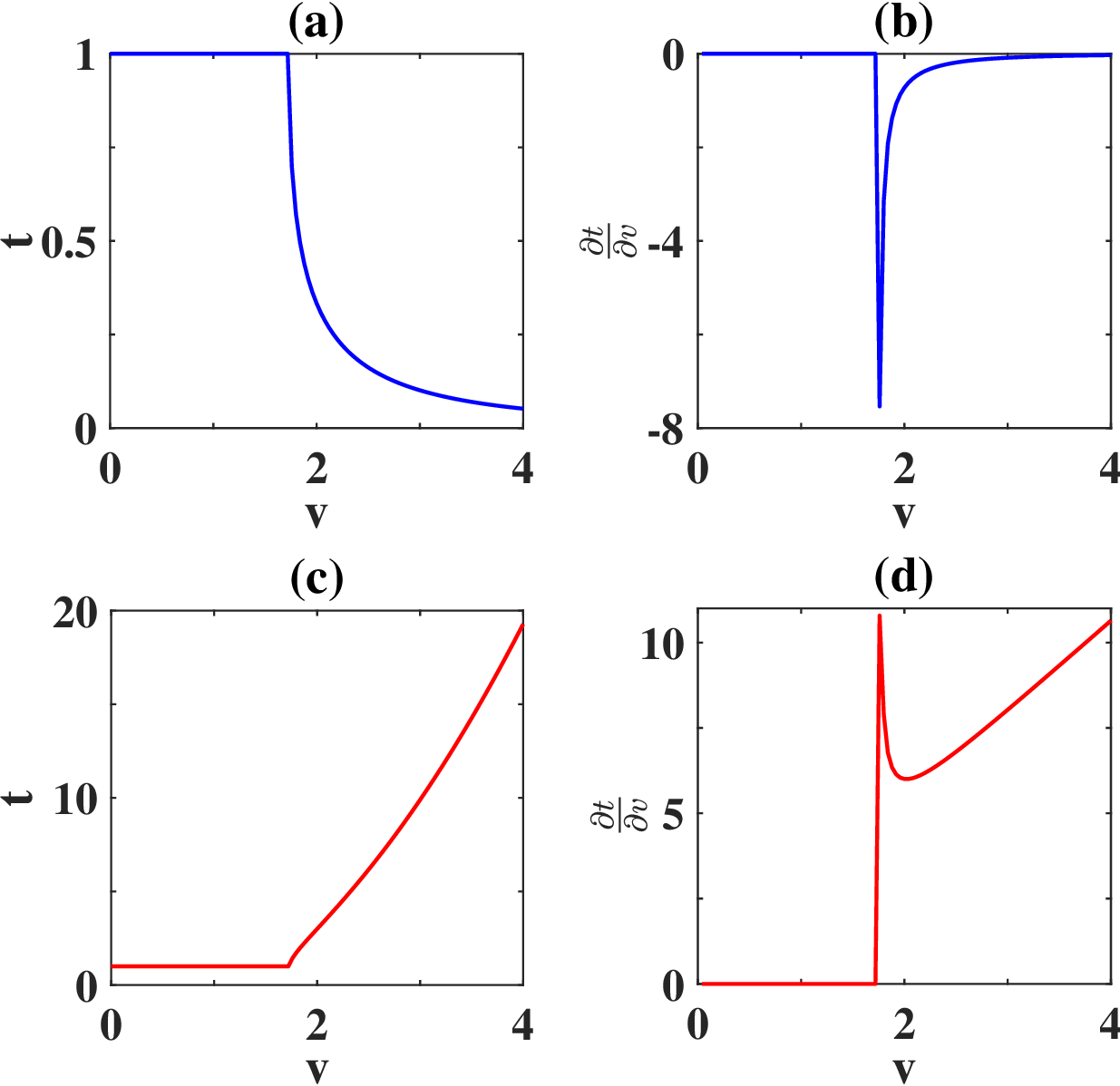}\caption{(a) and (b) are
the tearability $t$ and the derivative $\frac{\partial t}{\partial v}$ versus
the imaginary potential strength $v$ of the first boundary state,
respectively. (c) and (d) are the tearability $t$ and the derivative
$\frac{\partial t}{\partial v}$ versus as the imaginary potential strength $v$
of the fourth boundary state, respectively. Here, $t_{1}=1$ and $N=40$.}%
\end{figure}

According to Theorem \ref{thm1} in Eq. (\ref{eq1}), we give the effective Hamiltonian of bulk
states and boundary states, respectively. Firstly, the effective Hamiltonian
of bulk states can be written as
\begin{equation}
h_{\mathrm{eff}}\left(  k\right)  =\left(
\begin{array}
[c]{cc}%
h_{11} & h_{12}\\
h_{21} & h_{22}%
\end{array}
\right)
\end{equation}
with
\begin{align}
h_{11}  &  =h_{\mathrm{SSH}}\left(  k\right)  -ivI,\nonumber \\
h_{12}  &  =i\frac{\lambda}{\sqrt{N_{1}}}\left(
\begin{array}
[c]{cc}%
0 & t_{1}+t_{2}e^{-i\left(  k+\pi \right)  }\\
-\left[  t_{1}+t_{2}e^{i\left(  k+\pi \right)  }\right]  & 0
\end{array}
\right)  ,\nonumber \\
h_{21}  &  =h_{12},\nonumber \\
h_{22}  &  =h_{\mathrm{SSH}}\left(  k\right)  +ivI.
\end{align}
Here, $k=2\pi j/N_{1}$, $j=1,2,\cdots,N_{1}$, $N_{1}=N/2$, and $I=\left(
\begin{array}
[c]{cc}%
1 & 0\\
0 & 1
\end{array}
\right)  $. We present the complex energy spectrum from analytical solutions
of the effective Hamiltonian $H_{\mathrm{eff}}$ and
numerical solutions of the Hamiltonian $H=H_{\mathrm{SSH}}+V$ in Fig. 11. We
can see that they fit well, which manifests that the effective Hamiltonian we
give in the Theorem \ref{thm1} is feasible.
\begin{figure}[ptb]
\centering\includegraphics[clip,width=0.4\textwidth]{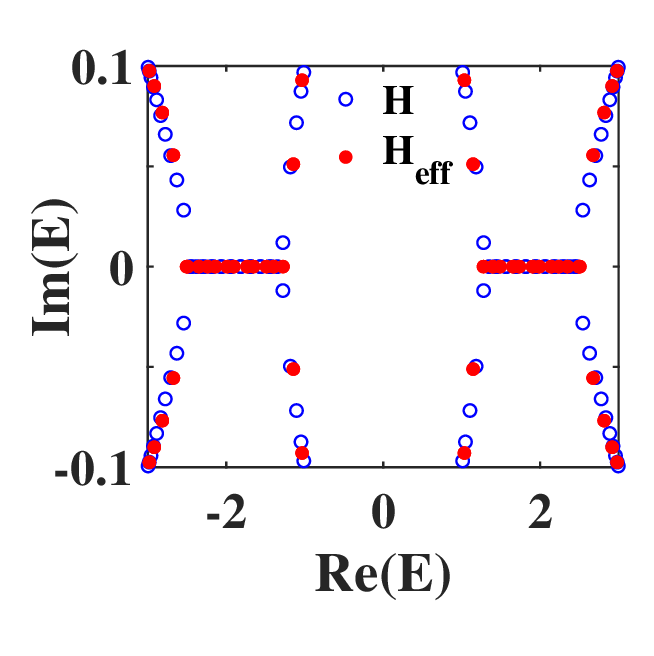}\caption{The complex energy
spectrum from numerical solutions of the Hamiltonian $H$ in the real space and
the analytical solutions of the effective Hamiltonian $H_{\mathrm{eff}}$ in
the $k$-space, where $v=0.1$ and $\lambda=0.392$. Here,
$t_{1}=1,t_{2}=2,$ and $N=50$. }%
\end{figure}

\begin{figure}[ptb]
\centering\includegraphics[clip,width=0.7\textwidth]{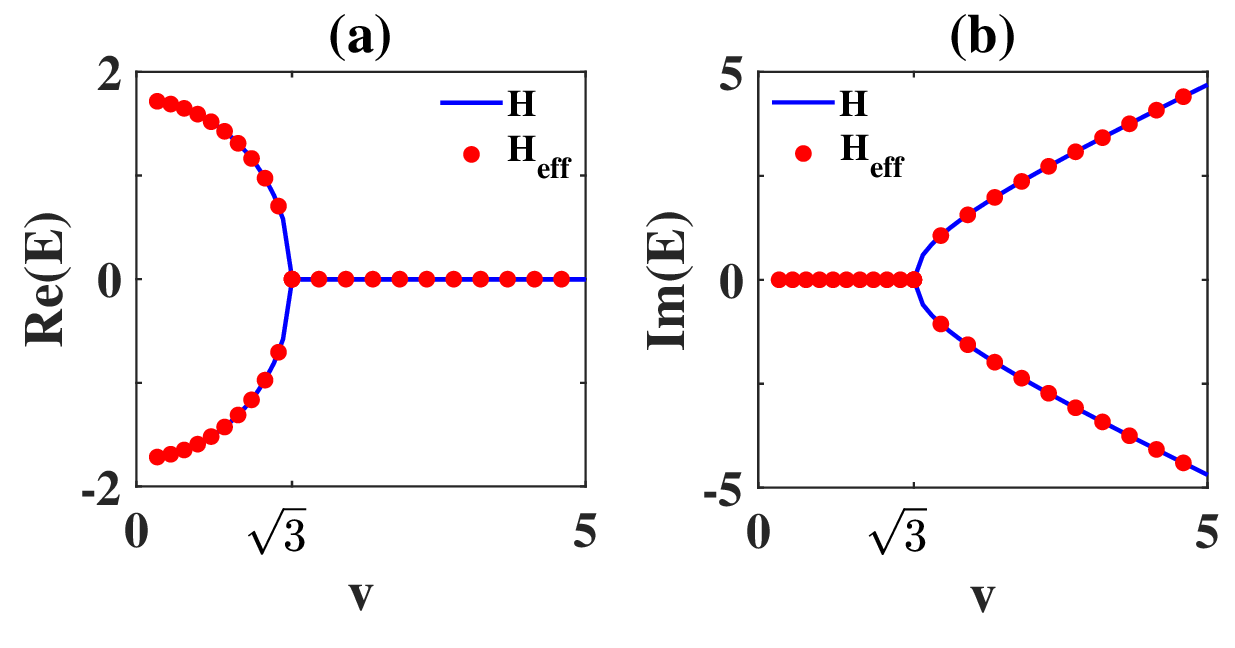}\caption{The numerical
solutions of boundary states from the Hamiltonian $H$ in the real space and
the analytical solutions of the effective Hamiltonian $H_{\mathrm{eff}}$ in
the $k$-space, respectively. Here, $t_{1}=1,t_{2}=2,$ and $N=50$. }%
\end{figure}
Secondly, the effective Hamiltonian of boundary states is expressed by
$H_{\mathrm{eff}}=I\otimes h_{\mathrm{eff}}$, where%
\begin{equation}
h_{\mathrm{eff}}=iv\sigma_{z}+\sqrt{3}\sigma_{y}.
\end{equation}
The eigenvalue is
\begin{equation}
E_{\mathrm{eff}}=\pm \sqrt{3-v^{2}}.
\end{equation}
Likewise, we show numerical solutions of $H=H_{\mathrm{SSH}}+V$ and analytical
solutions of $H_{\mathrm{eff}}$ for boundary states in
Fig. 12. We can see that $H_{\mathrm{eff}}$ agrees with
$H=H_{\mathrm{SSH}}+V$, which again confirms the rationality of Theorem \ref{thm1}.
There is a $\mathcal{PT}$ transition: in the case of $v<\sqrt{3}$, the
eigenvalues are all real and at $\mathcal{PT}$-symmetry; and in the case
of $v>\sqrt{3}$, the eigenvalues are all imaginary and at $\mathcal{PT}$-broken. At $v=\sqrt{3}$, boundary states are at the exceptional point with
energy degeneracy, i.e., $E_{\mathrm{eff}}\left(  k\right)  =0$.

\section{The Qi-Wu-Zhang model with the imaginary potential}\label{sec5}

In this section, we study the QWZ model with the imaginary potential to analyze
the properties of two-dimensional topological systems with the imaginary potential.

The Hamiltonian of the QWZ model in the real space is
\begin{align}
H_{\mathrm{QWZ}}  &  =\sum_{m_{x}=1}^{N_{x}-1}\sum_{m_{y}=1}^{N_{y}}\left(
|m_{x}+1,m_{y}\rangle \langle m_{x},m_{y}|\otimes t_{x}+h.c.\right) \nonumber \\
&  +\sum_{m_{x}=1}^{N_{x}}\sum_{m_{y}=1}^{N_{y}-1}\left(  |m_{x}%
,m_{y}+1\rangle \langle m_{x},m_{y}|\otimes t_{y}+h.c.\right) \nonumber \\
&  +u\sum_{m_{x}=1}^{N_{x}}\sum_{m_{y}=1}^{N_{y}}|m_{x},m_{y}\rangle \langle
m_{x},m_{y}|\otimes \sigma_{z},
\end{align}
where $u$ is the staggered on site potential. The model describes a particle
with two internal states hopping on a lattice where the nearest neighbour
hopping is accompanied by an operation on the internal degree of freedom, and
this operation is different for the hopping along the x-direction with $t_{x}%
=\frac{\sigma_{z}+i\sigma_{x}}{2}$ and y-direction with $t_{y}=\frac
{\sigma_{z}+i\sigma_{y}}{2}$. The Hamiltonian in the $k$-space is
\begin{equation}
h_{\mathrm{QWZ}}\left(  k\right)  =\sin k_{x}\cdot \sigma_{x}+\sin k_{y}%
\cdot \sigma_{y}+\left(  \cos k_{x}+\cos k_{y}+u\right)  \cdot \sigma_{z},
\end{equation}
and its eigenvalue is
\begin{equation}
E_{\pm}\left(  k\right)  =\pm \sqrt{\left(  \sin k_{x}\right)  ^{2}+\left(
\sin k_{y}\right)  ^{2}+\left(  \cos k_{x}+\cos k_{y}+u\right)  ^{2}}.
\end{equation}
We consider the imaginary potential%
\begin{align}
V  &  =-iv\sum_{m_{x}=1}^{L/2}\sum_{m_{y}=1}^{N/2}|m_{x},m_{y}\rangle \langle
m_{x},m_{y}|\otimes I\nonumber \\
&  +iv\sum_{m_{x}=L/2+1}^{L}\sum_{m_{y}=N/2+1}^{N}|m_{x},m_{y}\rangle \langle
m_{x},m_{y}|\otimes I
\end{align}
for the QWZ model. The potential $-iv$ is applied to the left $N/2\times L/2$
sites and the potential $iv$ is applied to the right $N/2\times L/2$ sites.

\begin{figure}[ptb]
\centering\includegraphics[clip,width=0.75\textwidth]{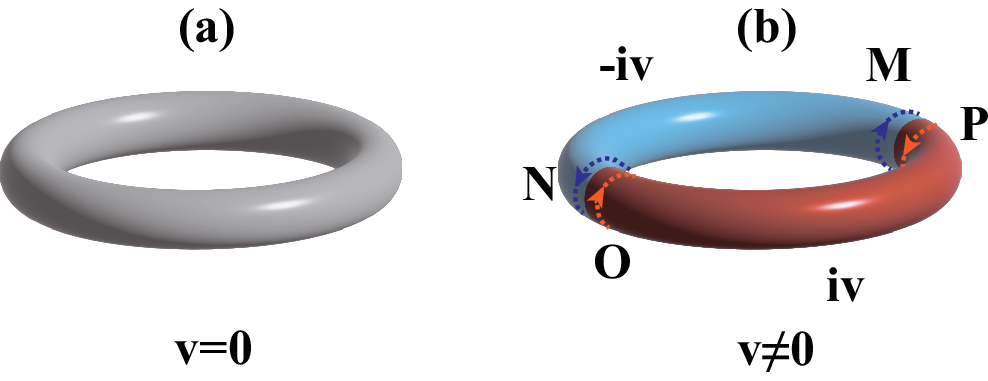}\caption{The diagram of the
Qi-Wu-Zhang model under the periodic conditions along the x-direction and
y-direction. (a) represents the case without the imaginary potential. (b)
represents the case with the imaginary potential. M, N, O, and P denote two
sides of two interfaces, respectively. The arrow represents the direction of the wave vector of boundary states.}%
\end{figure}

We consider the periodic boundary condition both along the
x-direction and the y-direction, as plotted in Fig. 13. The complex energy
spectra and eigenstates of the model are shown in Figs. 14 and 15. With an
augment of the imaginary potential strength $v$, bulk states show separation
within the $\mathcal{PT}$ transition. After the separation of bulk states, the
system appears some boundary states as reflected in Fig. 15. Each energy
eigenvalue of the boundary state has two eigenstates, so here we only take one
of the eigenstates in Fig. 15.
\begin{figure}[ptb]
\centering\includegraphics[clip,width=0.8\textwidth]{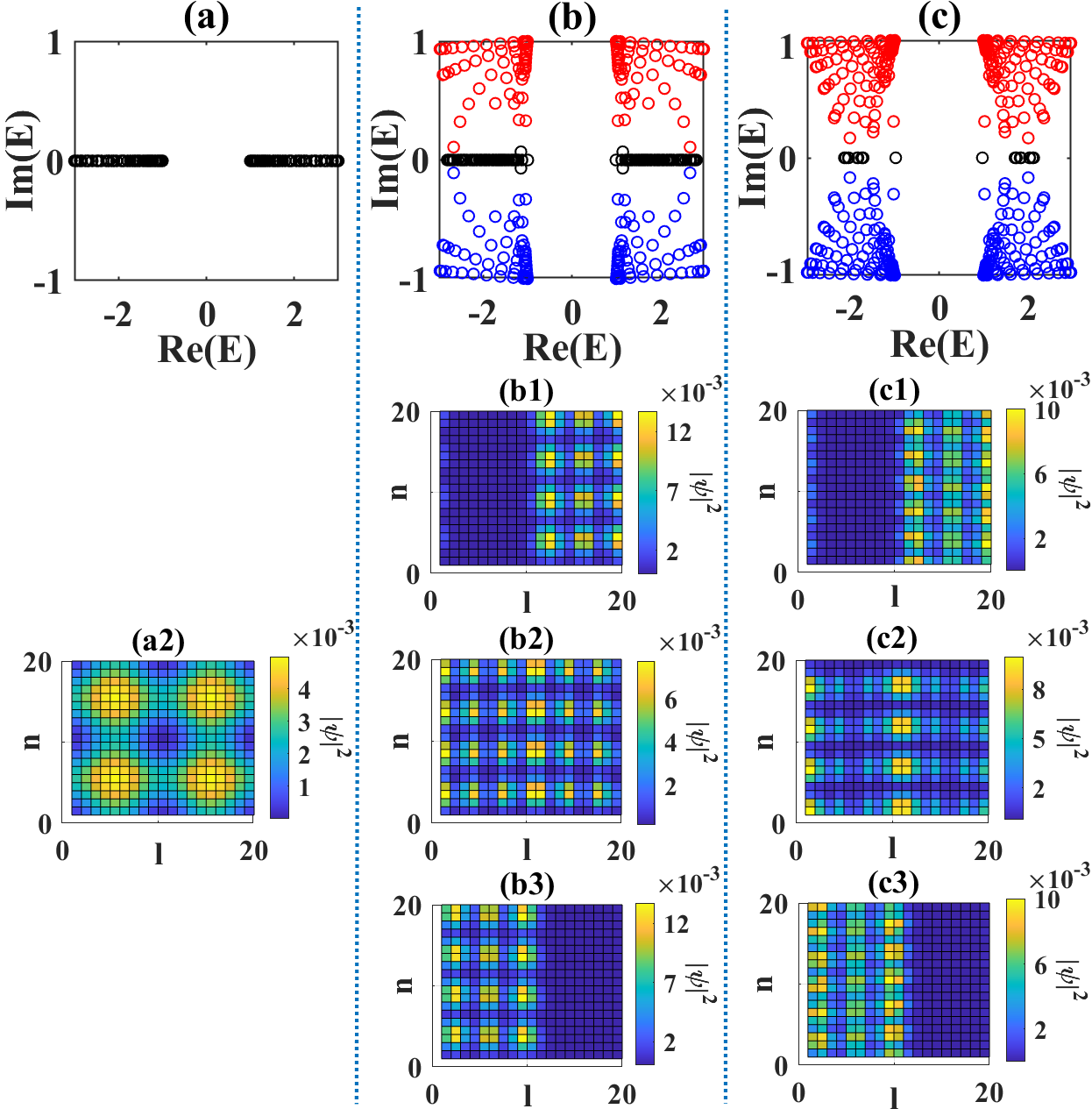}\caption{(a), (b) and (c) are the complex energy spectra for different imaginary potential strengths $v=0$, $0.1$ and $0.21$, respectively. (a2) is a
representative bulk state versus the site $n$ and $l$ for $v=0$. (b1) to (b3)
is a representative bulk state versus the site $n$ and $l$ corresponding to
the energy eigenvalues represented by the red, black, and blue circles for
$v=0.1$, respectively. (c1) to (c3) is a representative bulk state versus the
site $n$ and $l$ corresponding to the energy eigenvalues represented by the
red, black, and blue circles for $v=0.21$, respectively. Here, $N=L=20$ and
$u=-1$.}%
\end{figure}
\begin{figure}[ptb]
\centering\includegraphics[clip,width=0.8\textwidth]{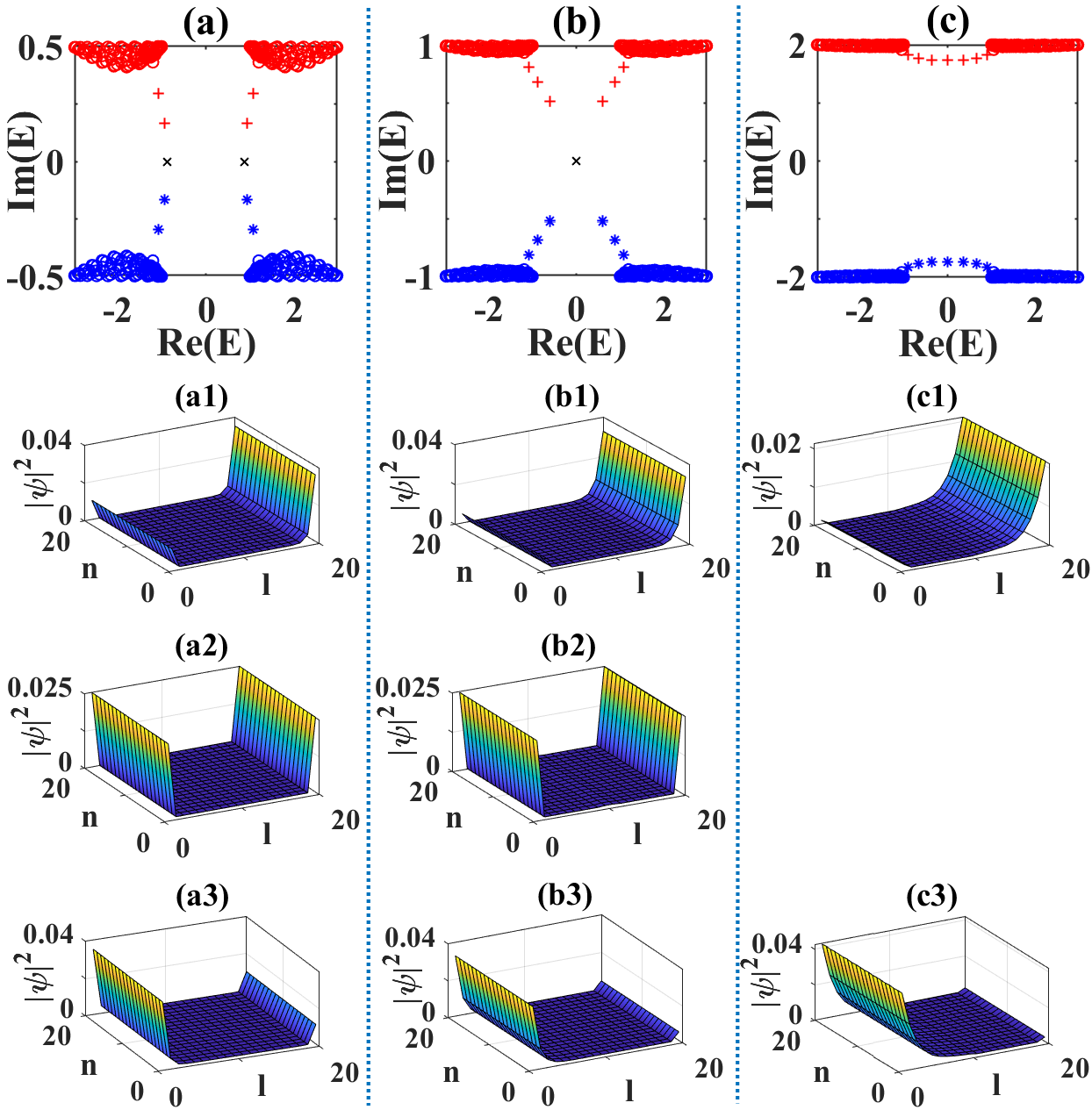}\caption{(a), (b) and (c) are the complex energy spectra for different
imaginary potential strengths $v=0.5$, $1$ and $2$, respectively. (a1) to (a3)
is a representative boundary state versus the site $n$ and $l$ corresponding
to the energy eigenvalues represented by the red plus signs, black crosses,
and blue asterisks for $v=0.5$, respectively. (b1) to (b3) is a representative
boundary state versus the site $n$ and $l$ corresponding to the energy
eigenvalues represented by the red plus signs, black crosses, and blue
asterisks for $v=1$, respectively. (c1) and (c3) are the representative
boundary states versus the site $n$ and $l$ corresponding to the energy
eigenvalues represented by the red plus signs and blue asterisks for $v=2$,
respectively. Here, $N=L=20$ and $u=-1$.}%
\end{figure}

As the imaginary potential strength $v$ continues to increase, these boundary
states show non-Hermitian tearing with a $\mathcal{PT}$ transition. As a
result, the wave functions of boundary states, corresponding to the down energy
band (represented by blue asterisks), are only localized at the side M or N of
the interface in Fig. 15(c3). The wave functions of boundary states,
corresponding to the up energy band (represented by red plus signs), are only
localized at the position O or P of the interface in Fig. 15(c1). The direction
of the wave vector of boundary states is parallel to the interfaces' direction, so
boundary states appear decoupling. Additionally, we calculate the tearability
and its derivative of the $j=350$ and the $j=450$ eigenstates in Fig. 16. There
is a second-order phase transition about the tearability at the exceptional
point.
\begin{figure}[ptb]
\centering\includegraphics[clip,width=0.7\textwidth]{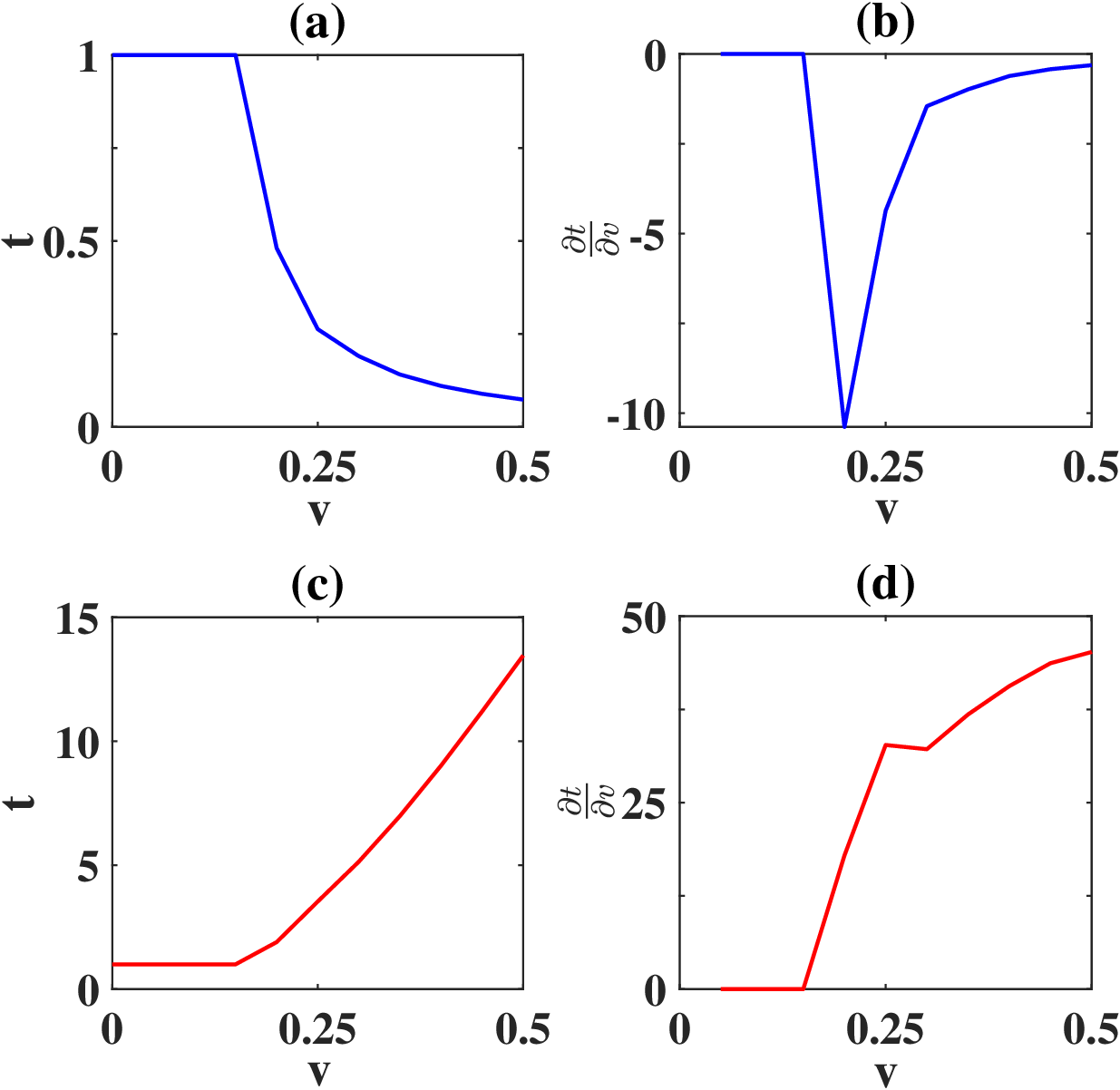}\caption{The tearability
$t$ and the derivative $\frac{\partial t}{\partial v}$ of the Qi-Wu-Zhang
model. (a) and (b) are $t$ and $\frac{\partial t}{\partial v}$ of the
$j=350$ eigenstate, respectively. (c) and (d) are $t$ and $\frac{\partial
t}{\partial v}$ of the $j=450$ eigenstate, respectively. Here, $N=L=20$ and
$u=-1$.}%
\end{figure}

Following Theorem \ref{thm1}, we provide the effective Hamiltonian for this
situation. The effective Hamiltonian of bulk states can be described by
\begin{equation}
h_{\mathrm{eff}}=\left(
\begin{array}
[c]{cc}%
h_{11} & h_{12}\\
h_{21} & h_{22}%
\end{array}
\right)  ,
\end{equation}
where%
\begin{align}
h_{11}  &  =h_{\mathrm{QWZ}}\left(  k\right)  -ivI,\nonumber \\
h_{12}  &  =\frac{\lambda_{1}}{\sqrt{L_{1}}}\left[  \sin \left(  k_{x}%
+\pi \right)  \cdot \sigma_{x}+\cos \left(  k_{x}+\frac{\pi}{2}\right)
\cdot \sigma_{z}\right]  ,\nonumber \\
h_{21}  &  =h_{12},\\
h_{22}  &  =h_{\mathrm{QWZ}}\left(  k\right)  +ivI.
\end{align}
$\lambda_{1}$ is the fitting parameter. Here, $k_x=2\pi j_x/L_{1}$, $j_x=1,2,\cdots,L_{1},$ and $L_{1}=L/2$.
$k_y=2\pi j_y/N$, $j_y=1,2,\cdots,N$.
For the boundary states, the effective
Hamiltonian is%
\begin{equation}
h_{\mathrm{eff}}=\left(
\begin{array}
[c]{cc}%
-\sin k_{y}-iv & \lambda_{2}\\
\lambda_{2} & \sin k_{y}+iv
\end{array}
\right)  ,
\end{equation}
where $\lambda_{2}$ is a fitting parameter. The numerical solutions of the
Hamiltonian $H=H_{\mathrm{QWZ}}+V$ in the real space and
analytical solutions of the effective Hamiltonian $H_{\mathrm{eff}}$ in the $k$-space for bulk states and boundary states
are shown in Fig. 17. Roughly, the solutions of effective Hamiltonian
are in agreement with numerical results.

\begin{figure}[ptb]
\centering\includegraphics[clip,width=0.4\textwidth]{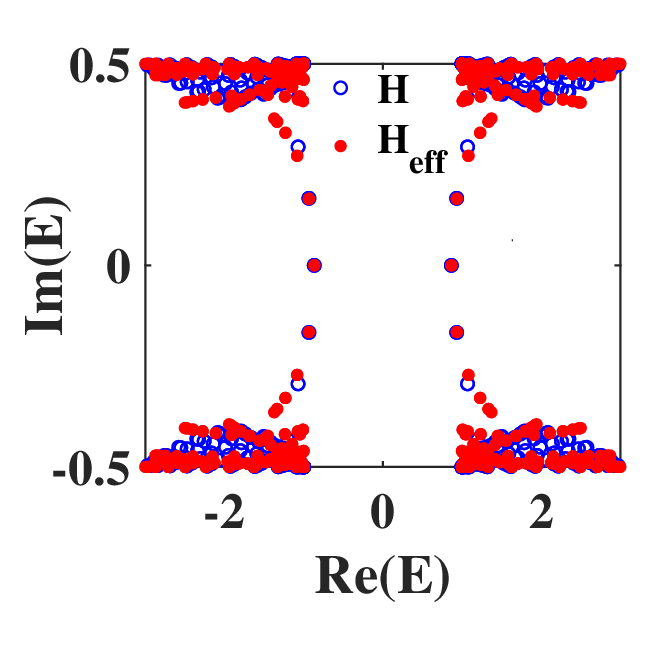}\caption{The complex energy
spectra from numerical solutions of the Hamiltonian $H$ in the real space and
solutions of the effective Hamiltonian $H_{\mathrm{eff}}$ in the $k$-space,
where $v=0.5$, $\lambda_{1}=0.75$ and $\lambda_{2}=0.9999$.}%
\end{figure}

For an arbitrary dimensional model $h_{0}\left(  k_{\bot},\vec{k}_{//}\right)$ with the imaginary potential, in which the
potential $-iv$ is applied to the left $L/2$ sites and the potential $iv$ is applied
to the right $L/2$ sites, the effective Hamiltonian of bulk states is
\begin{equation}
h_{\mathrm{eff}}\left(  k_{\bot},\vec{k}_{//}\right)  =\left(
\begin{array}
[c]{cc}%
h_{0}\left(  k_{\bot},\vec{k}_{//}\right)  -ivI & \frac{\alpha \left(  k_{\bot
},\vec{k}_{//}\right)  }{\sqrt{L/2}}\\
\frac{\alpha \left(  k_{\bot},\vec{k}_{//}\right)  }{\sqrt{L/2}} & h_{0}\left(
k_{\bot},\vec{k}_{//}\right)  +ivI
\end{array}
\right) ,
\end{equation}
where $k_{\bot}=2\pi j/L_{1}$, $j=1,2,\cdots,L_{1},$ and $L_{1}=L/2$. $I$ is a unit operator.
If this model is also topological, then it
would disappear boundary states after the bulk states show separation. The
effective Hamiltonian of these boundary states is%
\begin{equation}
h_{\mathrm{eff}}\left(  \vec{k}_{//}\right)  =\left(
\begin{array}
[c]{cc}%
h_{0}\left(  \vec{k}_{//}\right)  -ivI & \lambda I \\
\lambda I & -h_{0}\left(  \vec{k}_{//}\right)  +ivI
\end{array}
\right)  ,
\end{equation}
where $\lambda$ is a fitting parameter related to $v$.

\section{Conclusions}\label{sec6}

In the paper, we study the non-Hermitian system under dissipation, where the
left and right sites are subject to different imaginary potentials. To better
understand the physical phenomena of the total system, we give a series of the
effective $2\times2$ Hamiltonian in the $k$-space by reducing the $N\times N$
Hamiltonian in the real space. We discover that the energy band shows an
imaginary line gap with the imaginary potential. Based on this phenomenon, the theory of non-Hermitian tearing is proposed, in which the system is either in the partial tearing or complete tearing accompanied by a $\mathcal{PT}$
transition. To describe the effect of different imaginary potentials on an
eigenstate, we define tearability. According to the relationship between the
interface's direction and the direction of the wave vector of eigenstates,
separation and decoupling are introduced. By the theory of non-Hermitian
tearing, we explore the physical properties of the simple one-dimensional
tight binding model, one-dimensional SSH model and QWZ model. The results show
that in the process of non-Hermitian tearing, the system has the following
properties:
\begin{enumerate}[1.]
\item bulk states show separation;

\item after the separation of bulk states, topological systems could appear boundary states and these boundary states show decoupling;

\item the tearability exhibits a continuous phase transition at the exceptional point;

\item if the model adds the nonreciprocal hopping, then bulk states will display skin effect in the bound region, which will be our next study.
\end{enumerate}
In the future, we will further study the non-Hermitian tearing in
more complex systems and the relationship with other non-Hermitian phenomena.

\bmhead{Acknowledgements}

This work was supported by NSFC (Grant No. 11974053, 12174030) and the National Key R$\&$D Program of China (Grant No.2023YFA1406704).

\bmhead{Author contribution statement}
Qian Du and Su-Peng Kou contributed to the analysis and writing of the manuscript. Qian Du and Xin-Ran Ma contributed to the calculation of the manuscript. Every author has discussed and reviewed the manuscript before submission.
\\
\textbf{Data Availability Statement}
This manuscript has no associated date or the date will not be deposited.
\begin{appendices}

\section{The probability and tearability of eigenstates of the effective Hamiltonian}\label{secA1}

For the simple one-dimensional tight binding model with the imaginary potential, $h_{\mathrm{eff}}\left(  k\right)  $ is expressed as
\begin{equation}
h_{\mathrm{eff}}\left(  k\right)  =\left(
\begin{array}
[c]{cc}%
E_{0}-iv & \alpha \\
\alpha & E_{0}+iv
\end{array}
\right)  ,
\end{equation}
where $\alpha=\frac{2t\sin k}{\sqrt{N_{1}}}\cdot \lambda$ and $\lambda$ is a
fitting parameter related to $v$. The eigenvalues are
\begin{equation}
E_{\mathrm{eff}}=E_{0}\pm \sqrt{\alpha^{2}-v^{2}}.
\end{equation}
The eigenstates are
\begin{equation}
\Psi_{+}=\frac{1}{A}\binom{1}{\frac{iv+\sqrt{\alpha^{2}-v^{2}}}{\alpha}}%
,\Psi_{-}=\frac{1}{B}\binom{1}{\frac{iv-\sqrt{\alpha^{2}-v^{2}}}{\alpha}},
\end{equation}
where $A=\sqrt{1+\left(  \frac{iv+\sqrt{\alpha^{2}-v^{2}}}{\alpha}\right)
^{2}}$ and $B=\sqrt{1+\left(  \frac{iv-\sqrt{\alpha^{2}-v^{2}}}{\alpha
}\right)  ^{2}}$. The corresponding probabilities are
\begin{align}
\rho_{+,\mathrm{L}}  &  =\frac{\alpha^{2}}{\alpha^{2}+\left(  iv+\sqrt
{\alpha^{2}-v^{2}}\right)  ^{2}},\nonumber \\
\rho_{+,\mathrm{R}}  &  =\frac{\left(  iv+\sqrt{\alpha^{2}-v^{2}}\right)
^{2}}{\alpha^{2}+\left(  iv+\sqrt{\alpha^{2}-v^{2}}\right)  ^{2}},\nonumber \\
\rho_{-,\mathrm{L}}  &  =\frac{\alpha^{2}}{\alpha^{2}+\left(  iv-\sqrt
{\alpha^{2}-v^{2}}\right)  ^{2}},\nonumber \\
\rho_{-,\mathrm{R}}  &  =\frac{\left(  iv-\sqrt{\alpha^{2}-v^{2}}\right)
^{2}}{\alpha^{2}+\left(  iv-\sqrt{\alpha^{2}-v^{2}}\right)  ^{2}}.
\end{align}
The tearability are
\begin{align}
t_{+}  &  =\frac{\left(  iv+\sqrt{\alpha^{2}-v^{2}}\right)  ^{2}}{\alpha^{2}%
},\nonumber \\
t_{-}  &  =\frac{\left(  iv-\sqrt{\alpha^{2}-v^{2}}\right)  ^{2}}{\alpha^{2}}.
\end{align}

\end{appendices}


\begin{thebibliography}{99}                                                                                               %
\bibliographystyle{abbrv}

\makeatletter

\renewcommand\@biblabel[1]{#1.}

\makeatother


\bibitem {C2010}C. E. R\"{u}er, K. G. Makris, R. El-Ganainy, D. N.
Christodoulides, M. Segev, and D. Kip, Observation of parity--time symmetry in
optics, Nat. Phys. \textbf{6}, 192 (2010)

\bibitem {A2009}A. Guo, G. J. Salamo, D. Duchesne, R. Morandotti, M.
Volatier-Ravat, V. Aimez, G. A. Siviloglou, and D. N. Christodoulides,
Observation of $\mathcal{PT}$-Symmetry Breaking in Complex Optical Potentials
Phys. Rev. Lett. \textbf{103}, 093902 (2009)

\bibitem {Y2011}Y. D. Chong, L. Ge, and A. D. Stone, $\mathcal{PT}$-Symmetry
Breaking and Laser-Absorber Modes in Optical Scattering Systems, Phys. Rev.
Lett. \textbf{106}, 093902 (2011)

\bibitem {R2018}R. El-Ganainy, K. G. Makris, M. Khajavikhan, Z. H. Musslimani,
S. Rotter, and D. N. Christodoulides, Non-Hermitian physics and PT symmetry,
Nat. Phys. \textbf{14}, 11 (2018)

\bibitem {L2013}L. Feng, Y.-L. Xu, W. S. Fegadolli, M.-H. Lu, J. E. B.
Oliveira, V. R. Almeida, Y.-F. Chen, and A. Scherer, Experimental
demonstration of a unidirectional reflectionless parity-time metamaterial at
optical frequencies, Nat. Mater. \textbf{12}, 108 (2013)

\bibitem {H2014}H. Hodaei, M.-A. Miri, M. Heinrich, D. N. Christodoulides, and
M. Khajavikhan, Parity-time-symmetric microring lasers, Science \textbf{346},
975 (2014)

\bibitem {L2014}L. Feng, Z. J. Wong, R.-M. Ma, Y. Wang, and X. Zhang, Single
mode laser by parity-time symmetry breaking, Science \textbf{346}, 972 (2014)

\bibitem {W2019}W. Song, W. Sun, C. Chen, Q. Song, S. Xiao, S. Zhu, and T. Li,
Breakup and Recovery of Topological Zero Modes in Finite Non-Hermitian Optical
Lattices, Phys. Rev. Lett. \textbf{123}, 165701 (2019)

\bibitem {V2017}V. Kozii and L. Fu, Non-Hermitian Topological Theory of
Finite-Lifetime Quasiparticles: Prediction of Bulk Fermi Arc due to
Exceptional Point, arXiv:1708.05841

\bibitem {Z2018}Z. Gong, Y. Ashida, K. Kawabata, K. Takasan, S. Higashikawa,
and M. Ueda, Topological Phases of Non Hermitian Systems, Phys. Rev. X
\textbf{8}, 031079 (2018)

\bibitem {H2018}H. Shen, B. Zhen, and L. Fu, Topological Band Theory for
Non-Hermitian Hamiltonians, Phys. Rev. Lett. \textbf{120}, 146402 (2018)

\bibitem {F2019}F. Song, S. Yao, and Z. Wang, Non-Hermitian Topological
Invariants in Real Space, Phys. Rev. Lett. \textbf{123}, 246801 (2019)

\bibitem {M2020}N. Matsumoto, K. Kawabata, Y. Ashida, S. Furukawa, and M.
Ueda, Continuous Phase Transition without Gap Closing in Non-Hermitian Quantum
Many-Body Systems, Phys. Rev. Lett. \textbf{125}, 260601 (2020)

\bibitem {K2019}K. Kawabata, K. Shiozaki, M. Ueda, and M. Sato, Symmetry and
Topology in Non-Hermitian Physics, Phys. Rev. X \textbf{9}, 041015 (2019)

\bibitem {K2021}K. Kawabata, K. Shiozaki, and S. Ryu, Topological Field Theory
of Non-Hermitian Systems, Phys. Rev. Lett. \textbf{126}, 216405 (2021)

\bibitem {Y2020}Y. Michishita and R. Peters, Equivalence of Effective
Non-Hermitian Hamiltonians in the Context of Open Quantum Systems and Strongly
Correlated Electron Systems, Phys. Rev. Lett. \textbf{124}, 196401 (2020)

\bibitem {Bender02}C. M. Bender, D. C. Brody, and H. F. Jones, Complex
Extension of Quantum Mechanics, Phys. Rev. Lett. \textbf{89}, 270401 (2002)

\bibitem {I2009}I. Rotter, A non-Hermitian Hamilton operator and the physics
of open quantum systems, J. Phys. A: Math. Theor. \textbf{42}, 153001 (2009)

\bibitem {F2012}F. Reiter and A. S. S{\o }rensen, Effective operator formalism
for open quantum systems, Phys. Rev. A \textbf{85}, 032111 (2012)

\bibitem {M2002}A. Mostafazadeh, Pseudo-Hermiticity versus PT symmetry: The
necessary condition for the reality of the spectrum of a non-Hermitian
Hamiltonian J. Math. Phys. \textbf{43 }205 (2002); A. Mostafazadeh,
Pseudo-Hermiticity versus PT-symmetry. II. A complete characterization of
non-Hermitian Hamiltonians with a real spectrum, ibid, \textbf{43} 2814
(2002); A. Mostafazadeh A, Pseudo-Hermiticity versus PT-symmetry III:
Equivalence of pseudo-Hermiticity and the presence of antilinear symmetries,
ibid, \textbf{43} 3944, (2002)

\bibitem {N2011}N. Moiseyev, Non-Hermitian Quantum Mechanics (Cambridge
University Press, Cambridge, 2011)

\bibitem {U2020}Y. Ashida, Z. Gong, and M. Ueda, Non-Hermitian physics, Adv.
Phys. \textbf{69}, 249 (2020)

\bibitem {Bender07}C. M. Bender, Making sense of non-Hermitian Hamiltonians,
Rep. Prog. Phys. \textbf{70}, 947 (2007).

\bibitem {G1928}G. Gamow, Zur Quantentheorie des Atomkernes, Z. Phys.
\textbf{51}, 204 (1928)

\bibitem {P1982}P. M. Radmore and P. L. Knight, Population trapping and
dispersion in a three-level system, J. Phys. B: Atom. Mol. Phys., \textbf{15},
561 (1982)

\bibitem {R1971}R. M. More, Theory of decaying states, Phys. Rev. A
\textbf{4}, 1782 (1971)

\bibitem {H1996}N. Hatano and D. R. Nelson, Localization Transitions in
Non-Hermitian Quantum Mechanics, Phys. Rev. Lett. \textbf{77}, 570 (1996)

\bibitem {Bender98}C. M. Bender, and S. Boettcher, Phys. Rev. Lett.
\textbf{80}, 5243 (1998)

\bibitem {V2018}V. M. Martinez Alvarez, J. E. Barrios Vargas, and L. E. F. Foa
Torres, Non-Hermitian robust edge states in one dimension: Anomalous
localization and eigenspace condensation at exceptional points, Phys. Rev. B
\textbf{97}, 121401(R) (2018)

\bibitem {Y2018}S. Yao and Z. Wang, Edge States and Topological Invariants of
Non-Hermitian Systems, Phys. Rev. Lett. \textbf{121}, 086803 (2018)

\bibitem {L2020}L. Li, C. H. Lee, S. Mu, and J. Gong, Critical non-Hermitian
skin effect, Nat. Commun. \textbf{11}, 5491 (2020)

\bibitem {K2020}K. Kawabata, M. Sato, and K. Shiozaki, Higher-order
non-hermitian skin effect, Phys. Rev. B \textbf{102}, 205118 (2020)

\bibitem {D2020}D. S. Borgnia, A. J. Kruchkov, and R.-J. Slager, Non-Hermitian
Boundary Modes and Topology, Phys. Rev. Lett. \textbf{124}, 056802 (2020)

\bibitem {N2020}N. Okuma, K. Kawabata, K. Shiozaki, and M. Sato, Topological
Origin of Non-Hermitian Skin Effects, Phys. Rev. Lett. \textbf{124}, 086801 (2020)

\bibitem {X2018}Y. Xiong, Why does bulk boundary correspondence fail in some
non-hermitian topological models, J. Phys. Commun. \textbf{2} 035043 (2018)

\bibitem {T2016}T. E. Lee, Anomalous Edge State in a Non-Hermitian Lattice,
Phys. Rev. Lett. \textbf{116}, 133903 (2016)

\bibitem {F2018}F. K. Kunst, E. Edvardsson, J. C. Budich, and Emil J.
Bergholtz, Biorthogonal Bulk-Boundary Correspon dence in Non-Hermitian
Systems, Phys. Rev. Lett. \textbf{121}, 026808 (2018)

\bibitem {S2018}S. Yao, F. Song and Z. Wang, Non-Hermitian Chern Bands, Phys.
Rev. Lett. \textbf{121}, 136802 (2018)

\bibitem {H2021}H.-G. Zirnstein, G. Refael, and B. Rosenow, Bulk Boundary
Correspondence for Non-Hermitian Hamilto nians via Green Functions, Phys. Rev.
Lett. \textbf{126}, 216407 (2021)

\bibitem {X2020}X.-R. Wang, C.-X. Guo, and S.-P. Kou, Defective edge states
and number-anomalous bulk-boundary correspondence in non-Hermitian topological
systems, Phys. Rev. B \textbf{101}, 121116(R) (2020)

\bibitem {H2019}H. Zhao, X. Qiao, T. Wu, B. Midya, S. Longhi, and L. Feng,
Non-Hermitian topological light steering, Science \textbf{365}, 1163 (2019)

\bibitem {Y2022}Y. Li, C. Fan, X.g Hu, Y. Ao, C. Lu, C. T. Chan, D. M. Kennes,
and Q. Gong, Effective Hamiltonian for Photonic Topological Insulator with
Non-Hermitian Domain Walls, Phys. Rev. Lett. \textbf{129}, 053903 (2022)

\bibitem {J2022}Y.-J. Wu, C.-C. Liu and J. Hou, Wannier-type photonic
higher-order topological corner states induced solely by gain and loss, Phys.
Rev. A \textbf{101}, 043833 (2020)

\bibitem {D2023}D. Halder, S. Ganguly, and S. Basu, Properties of the
non-Hermitian SSH model: role of $\mathcal{PT}$ symmetry, J. Phys.: Condens.
Matter \textbf{35}, 105901 (2023)

\bibitem {S2023}S. Jana, and L. Sirota, Emerging exceptional point with
breakdown of skin effect in non-Hermitian systems, arXiv:2303.15050v2 (2023)

\bibitem {T2019}T.-S. Deng and W. Yi, Non-Bloch topological invariants in a
non-Hermitian domain wall system, Phys. Rev. B \textbf{100}, 035102 (2019)

\bibitem {J2023}J.-R. Li, C. Luo, L.-L. Zhang, S.-F. Zhang, P.-P. Zhu, and
W.-J. Gong, Band structures and skin effects of coupled nonreciprocal
Su-Schrieffer-Heeger lattices, Phys. Rev. A \textbf{107}, 022222 (2023)

\bibitem {C2023}C.-X. Guo, X. Wang, H. Hu, and S. Chen, Accumulation of
scale-free localized states induced by local non-Hermiticity, Phys. Rev. B
\textbf{107}, 134121 (2023)

\bibitem {B2022}B. Li, H.-R. Wang, F. Song and Z. Wang, Scale-free
localization and pt symme try breaking from local non-hermiticity,
arXiv:2302.04256v1 (2022)

\bibitem {X2023}X.-R. Ma, K. Cao, X.-R. Wang, Z. Wei, Q. Du, S.-P. Kou,
Non-Hermitian Chiral Skin Effect, Phys. Rev. Res. 6,013213(2024)

\end{thebibliography}
\end{document}